\documentclass[5p]{elsarticle}
\usepackage{graphicx}
\usepackage{amssymb}
\usepackage{amsmath}
\usepackage{multicol}
\usepackage{caption}
\usepackage{subcaption}

\journal{Astroparticle Physics}

\title{Improving Photoelectron Counting and Particle Identification in Scintillation Detectors with Bayesian Techniques}

\usepackage{xspace} 
\newcommand{\artn}{$^{39}\mathrm{Ar}$\xspace}
\newcommand{\fp}{$f_p$\xspace}
\newcommand{\rp}{$r_p$\xspace}
\newcommand{\lr}{$l_r$\xspace}
\newcommand{\na}{$^{22}$Na\xspace}

\begin{document}

\address[alberta]{Department of Physics, University of Alberta, Edmonton, Alberta, T6G 2R3, Canada}
\address[bu]{Department of Physics, Boston University, Boston, MA 02215, USA}
\address[ucb]{Department of Physics, University of California, Berkeley, Berkeley, CA 94720, USA}
\address[carl]{Department of Physics, Carleton University, Ottawa, Ontario, K1S 5B6, Canada}
\address[lu]{Department of Physics and Astronomy, Laurentian University, Sudbury, Ontario, P3E 2C6, Canada}
\address[lanl]{Los Alamos National Laboratory, Los Alamos, NM  87545, USA}
\address[mit]{Department of Physics, Massachusetts Institute of Technology, Cambridge, MA 02139, USA}
\address[nist]{National Institute of Standards and Technology, Boulder, CO 80305, USA}
\address[unm]{University of New Mexico, Albuquerque, NM 87131, USA}
\address[unc]{Department of Physics and Astronomy, University of North Carolina, Chapel Hill, NC 27599, USA}
\address[pnnl]{Pacific Northwest National Laboratory, Richland, WA  99352, USA}
\address[penn]{Department of Physics and Astronomy, University of Pennsylvania, Philadelphia, PA 19104, USA}
\address[queens]{Department of Physics, Engineering Physics, and Astronomy, Queen's University, Kingston, Ontario, K7L 3N6, Canada}
\address[rhul]{Department of Physics, Royal Holloway, University of London, Egham, Surrey TW20 0EX, UK}
\address[snolab]{SNOLAB Institute, Lively, ON P3Y 1N2, Canada}
\address[usd]{Department of Physics, University of South Dakota, Vermillion, SD 57069, USA}
\address[syr]{Physics Department, Syracuse University, Syracuse, NY 13244, USA}
\address[tunl]{Triangle Universities Nuclear Laboratory, Durham, NC 27708, USA}
\address[triumf]{TRIUMF, Vancouver, British Columbia, V6T 2A3, Canada}
\address[yale]{Department of Physics, Yale University, New Haven, CT 06520, USA}

\author[lanl]{M.~Akashi-Ronquest}
\author[triumf]{P.-A. Amaudruz}%
\author[lu,carl]{M.~Batygov}%
\author[alberta]{B.~Beltran}%
\author[unm]{M.~Bodmer}
\author[queens]{M.G.~Boulay}%
\author[queens]{B.~Broerman}%
\author[mit]{B.~Buck}
\author[rhul]{A.~Butcher}
\author[queens]{B.~Cai}%
\author[penn]{T.~Caldwell\corref{corauth}}
\cortext[corauth]{Corresponding author.}
\ead{tcald@hep.upenn.edu}
\author[queens]{M.~Chen}%
\author[syr]{Y.~Chen}
\author[snolab,lu]{B.~Cleveland}%
\author[nist]{K.~Coakley}
\author[queens]{K.~Dering}%
\author[snolab,lu]{F.A.~Duncan}
\author[mit]{J.A.~Formaggio}
\author[queens]{R.~Gagnon}%
\author[bu]{D.~Gastler}
\author[unm]{F.~Giuliani}
\author[unm]{M.~Gold}
\author[queens]{V.V.~Golovko}%
\author[alberta]{P.~Gorel}%
\author[carl]{K.~Graham}%
\author[rhul]{E.~Grace}
\author[mit]{N.~Guerrero}
\author[usd]{V.~Guiseppe}
\author[alberta]{A.L.~Hallin}%
\author[queens]{P.~Harvey}%
\author[queens]{C.~Hearns}%
\author[unc,tunl]{R.~Henning}
\author[lanl,pnnl]{A.~Hime}
\author[snolab]{J.~Hofgartner}%
\author[mit]{S.~Jaditz}
\author[snolab,lu]{C.J.~Jillings}%
\author[bu]{C.~Kachulis}
\author[bu]{E.~Kearns}
\author[mit]{J.~Kelsey}
\author[penn]{J.R.~Klein}
\author[queens]{M.~Ku\'zniak}%
\author[penn]{A.~LaTorre}
\author[snolab]{I.~Lawson}
\author[snolab]{O.~Li}
\author[queens]{J.J.~Lidgard}%
\author[snolab]{P.~Liimatainen}
\author[bu]{S.~Linden}
\author[carl]{K.~McFarlane}
\author[yale]{D.N.~McKinsey}
\author[unc,tunl]{S.~MacMullin}
\author[penn]{A.~Mastbaum}
\author[queens]{R.~Mathew}%
\author[queens]{A.B.~McDonald}%
\author[usd]{D.-M.~Mei}
\author[rhul]{J.~Monroe}
\author[triumf]{A. Muir}%
\author[snolab]{C.~Nantais}%
\author[queens]{K.~Nicolics}%
\author[rhul]{J.A.~Nikkel}
\author[queens]{T.~Noble}%
\author[queens]{E.~O'Dwyer}%
\author[alberta]{K.~Olsen}%
\author[ucb]{G.D.~Orebi Gann}
\author[carl]{C.~Ouellet}%
\author[snolab]{K.~Palladino}
\author[queens]{P.~Pasuthip}%
\author[usd]{G.~Perumpilly}
\author[lu]{T.~Pollmann}%
\author[queens]{P.~Rau}%
\author[triumf]{F. Reti\`ere}%
\author[lanl]{K.~Rielage}
\author[syr]{R.~Schnee}
\author[penn]{S.~Seibert}
\author[queens]{P.~Skensved}%
\author[queens]{T.~Sonley}%
\author[snolab]{E.~V\'azquez-J\'auregui}%
\author[queens]{L.~Veloce}
\author[rhul]{J.~Walding}
\author[syr]{B.~Wang}
\author[unm]{J.~Wang}
\author[queens]{M.~Ward}
\author[syr]{C.~Zhang}

\begin{abstract}
Many current and future dark matter and neutrino detectors are designed to measure scintillation light with a large array of photomultiplier tubes (PMTs).  The energy resolution and particle identification capabilities of these detectors depend in part on the ability to accurately identify individual photoelectrons in PMT waveforms despite large variability in pulse amplitudes and pulse pileup.  We describe a Bayesian technique that can identify the times of individual photoelectrons in a sampled PMT waveform without deconvolution, even when pileup is present.  To demonstrate the technique, we apply it to the general problem of particle identification in single-phase liquid argon dark matter detectors.  Using the output of the Bayesian photoelectron counting algorithm described in this paper, we construct several test statistics for rejection of backgrounds for dark matter searches in argon.  Compared to simpler methods based on either observed charge or peak finding, the photoelectron counting technique improves both energy resolution and particle identification of low energy events in calibration data from the DEAP-1 detector and simulation of the larger MiniCLEAN dark matter detector.
\end{abstract}


\maketitle

\section{Introduction}
\label{sec:intro}

Scintillators are a key component of many neutrino and dark matter experiments due to their relatively low cost and high light yield.  In addition, many scintillators can be used for particle identification when the time profile of the scintillation light is sensitive to the energy loss characteristics of different particles.  Experiments that require low energy thresholds, high energy resolution, and/or high levels of background rejection often combine a scintillating target medium with an array of photomultiplier tubes (PMTs) and a data acquisition system with sensitivity to individual photoelectrons.  In order to fully capture the scintillation time profile, the PMT pulses are typically recorded using a waveform digitizer.

Many common scintillators produce light on at least two characteristic time scales.    Particle identification in such scintillation detectors employs two related features:  the scintillation time scales, and the probability of populating the different time scales, which depends on the particle's energy loss characteristics.  The canonical approach (examples include MicroCLEAN \cite{microclean2008}, DEAP-1 \cite{deap1scint}, XMASS \cite{xenonpsd}, XENON-10 \cite{xenon10}, GERDA \cite{gerdapsd}, KamLAND \cite{kamlandpsd}) to time-based particle identification with a digitized time-dependent voltage waveform $V(t)$ is to estimate the fraction of the light produced on a fast timescale relative to the total amount of light produced in the event.  This particle discriminant we refer to as the prompt-fraction, or \fp, and is defined as
\begin{equation}
\label{eq:fp}
f_p = \frac{\int_{T_i}^\epsilon V(t) dt}{\int_{T_i}^{T_f} V(t) dt},
\end{equation}
where $T_i$ is some time before the prompt peak, $T_f$ is the time defined by the end of the event window, and $\epsilon$ depends on the timing characteristics of the scintillator.  Typically the \fp parameter is used to place a cut or perform some likelihood-based analysis in order to select a certain class of interactions in the scintillator.  In later sections, we refer to $f_p$ {\it leakage} as the probability of events from a certain class of background leaking into the \fp signal region of interest.  Although particle identification with \fp is robust in the sense that it is fairly insensitive to fluctuations in the scintillation light production or PMTs, its discrimination power breaks down at low energies beyond statistical effects because it loses information about the precise timing and charge of individual photoelectrons created by photons produced in the scintillator. 

\begin{figure}
\begin{center}
\includegraphics[width=1.0\columnwidth]{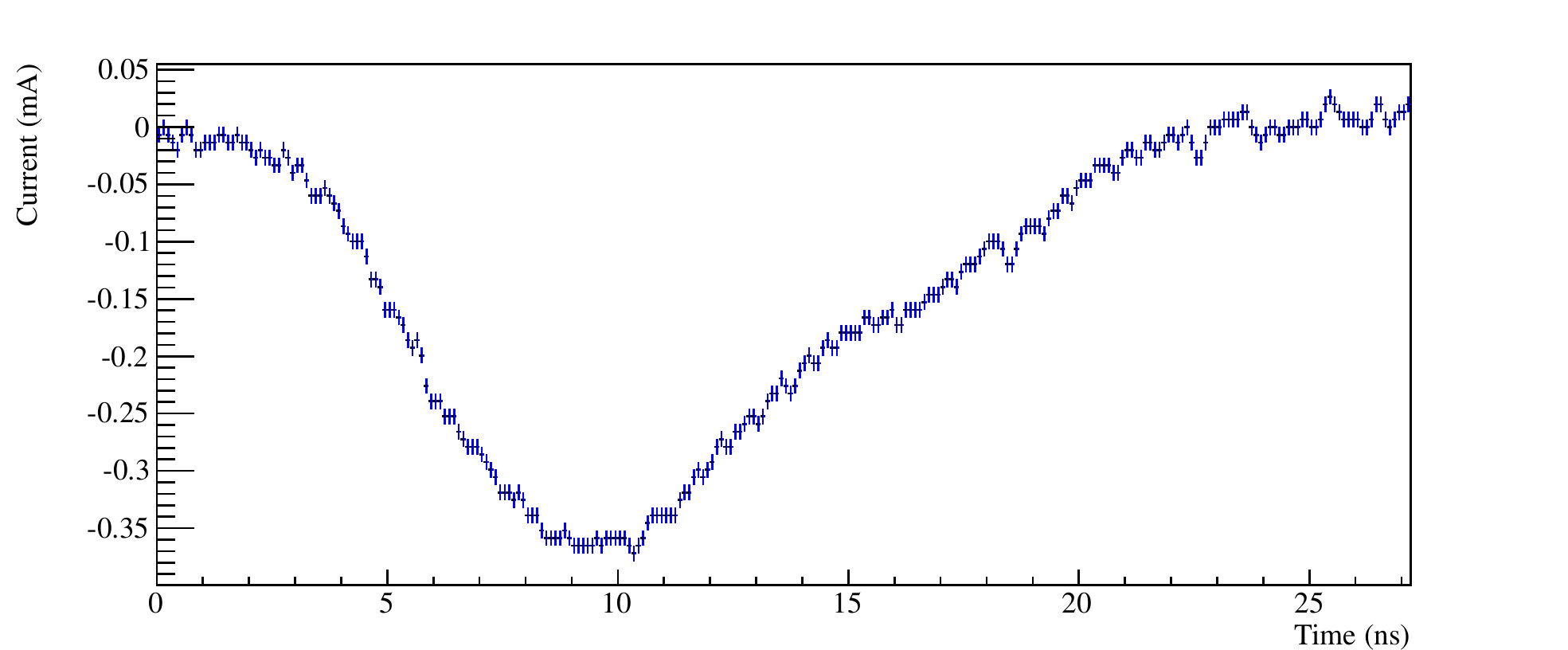}
\caption{\label{fig:pmt_pulse}Single photoelectron pulse from a Hamamatsu R5912-02-MOD PMT, as digitized by a 10 GHz oscilloscope.}
\end{center}
\end{figure}

\begin{figure}
\begin{center}
\includegraphics[width=1.0\columnwidth]{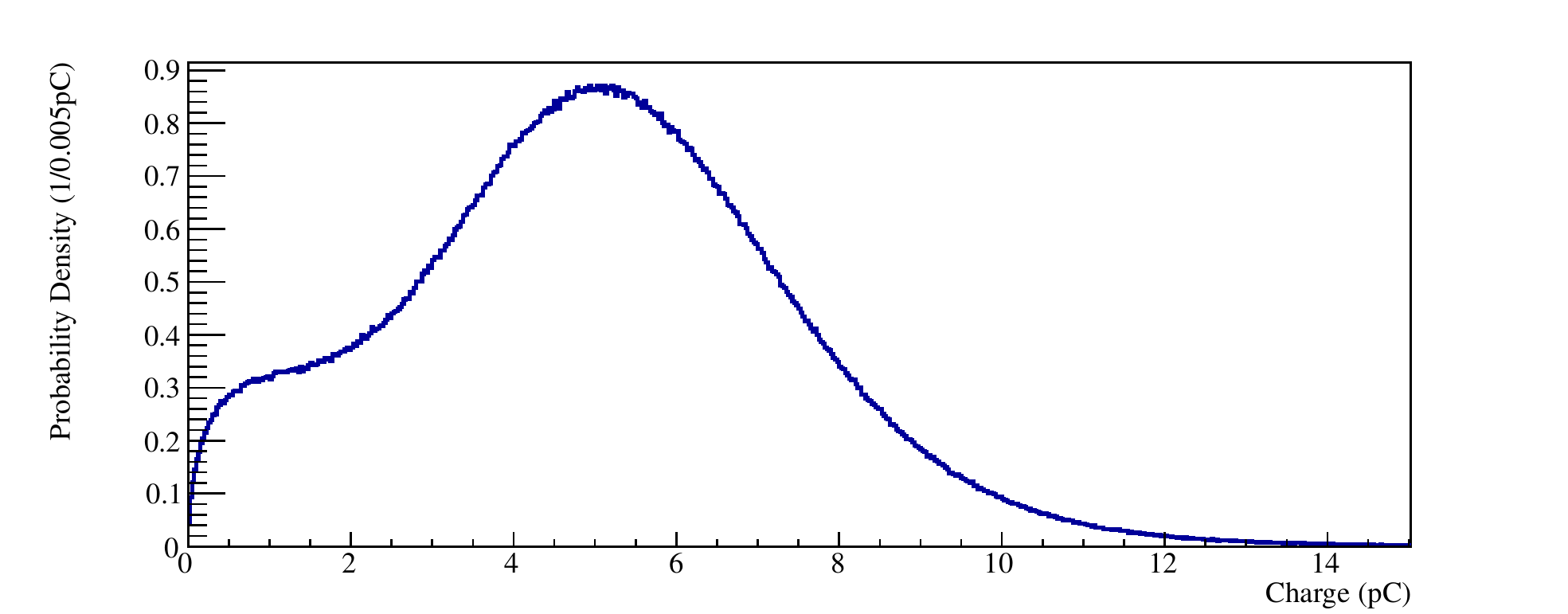}
\caption{\label{fig:charge_dist}The single photoelectron charge distribution of the Hamamatsu Photonics R5912-02-MOD cryogenic 8" PMT at a gain of $3\times 10^7$.  As described in~\cite{r591202mod}, the charge distribution is populated using only pulses which are adequate fits to a pulse shape model, so no assumptions about the shape of the `pedestal' near zero are required.}
\end{center}
\end{figure}

The best strategy to count photoelectron (PE) pulses in a waveform depends on the intensity and time structure of the light, as well as the characteristics of the PMT electronics.  A typical single photoelectron pulse from a large area PMT, shown for cryogenic measurements of a Hamamatsu Photonics R5912-02-MOD\footnote{Certain commercial equipment, instruments, or materials are identified in this report in order to specify the experimental procedure adequately. Such identification is not intended to imply recommendation or endorsement by the National Institute of Standards and Technology, nor is it intended to imply that the materials or equipment identified are necessarily the best available for the purpose.} 8" PMT in Figure~\ref{fig:pmt_pulse}, spans 20~ns or more~\cite{r591202mod}.  Additionally, there are large pulse-to-pulse variations in the amplification of a single photoelectron, resulting in a fairly broad charge distribution, shown again for the R5912-02-MOD in Figure~\ref{fig:charge_dist}.  If the light intensity observed by a PMT is very low or the time constant for scintillation light is very long (hundreds of nanoseconds or more) compared to the duration of a single photoelectron pulse, then overlap of pulses is improbable.  Photoelectrons can be counted by searching for peaks in the waveform, thus eliminating the impact of the PMT charge distribution on the counting procedure.

However, if multiple photoelectrons are likely to be detected by a PMT in a time much shorter than the single photoelectron pulse duration, such as from Cherenkov or fast scintillation light, peak finding is a poor photoelectron counting strategy.  Peaks from different photoelectrons may not be clearly resolved in the waveform, resulting in a systematic bias toward under-counting.  A simple and unbiased technique for photoelectron counting in this case would be to integrate the waveform and divide by the mean charge of a single photoelectron.  Due to the broad charge distribution of most PMTs, this normalized integral charge procedure in a waveform with pulse pileup will have more variance than peak counting in a waveform without pileup.  Fundamentally, information is lost in the pileup that is difficult to recover.

Realistic detectors typically span both of these extreme cases.  Many experiments observe both Cherenkov or fast scintillation \emph{and} slow scintillation light.  Moreover, the light intensity observed by a given PMT can vary dramatically depending on the energy of the event and its location in the detector.  To cover all these cases, we have designed a photoelectron counting method that combines both peak finding and normalized charge integration by using Bayes' Theorem to incorporate our external knowledge of how likely photoelectron pileup is for different events, PMTs, and times in the waveform.  The method also outputs an estimated photoelectron production time for each pulse, which can be analyzed for particle identification purposes in many scintillators.

To make the discussion concrete, we focus the photoelectron identification procedure specifically on liquid argon scintillation light, primarily in the MiniCLEAN dark matter experiment, but the general approach can be easily adapted to the constraints of other experiments.  In Section~\ref{sec:fp_model} we use a GPU-based fast Monte Carlo simulation of the MiniCLEAN detector to motivate the need for improvement in the canonical prompt-fraction particle identification technique.  The Bayesian photoelectron counting method is described in Section~\ref{sec:spe_counting}, and improved particle identification test statistics are defined in Section~\ref{sec:stat_tests}.  We then demonstrate the effectiveness of the Bayesian photoelectron counting method in Section~\ref{sec:results} with gamma simulation and calibration data collected from the DEAP-1 detector underground at SNOLAB.  In Section~\ref{sec:miniclean}, we describe the MiniCLEAN dark matter experiment and motivate the need for improved particle identification specifically in large liquid argon dark matter detectors.  Finally, Section~\ref{sec:sim} applies the Bayesian techniques to a complete Monte Carlo simulation of MiniCLEAN.

\section{Single-Phase Dark Matter Detection with Liquid Argon}

Laboratory searches for dark matter in the form of weakly-interacting massive particles (WIMPs) require sensitivity to nuclear recoils with tens of keV of kinetic energy.  WIMP-induced recoils need to be distinguished from other sources of low energy particle tracks, such as fast neutrons and radioactive decay in detector materials.  Dark matter experiments must simultaneously achieve very low levels of natural radioactivity and very high rejection factors for remaining sources of backgrounds.

Liquid argon offers a promising target for WIMP detection due to its combination of extremely low cost, moderate cryogenic requirements (similar to liquid nitrogen), straightforward purification techniques, and a reasonably high density and atomic number.  Argon is a very efficient scintillator, producing approximately 40 UV photons per keV~\cite{argonly} of deposited energy from an electron track.   Although very low levels of chemical impurities can be achieved with liquid argon, atmospheric argon contains 1~Bq/kg of \artn~\cite{ar39_activity}, a beta decay isotope with an endpoint energy of 565~keV and half-life of 269~years.  The \artn background must be mitigated either through the acquisition of argon depleted of \artn, through highly efficient identification and rejection of electronic recoils in the data, or both.

``Single-phase'' liquid argon experiments search for nuclear recoil events using only scintillation light, in the absence of an electric field~\cite{boulayhime}. Prototype detectors, such as MicroCLEAN~\cite{microclean2008,microclean2012} and DEAP-1~\cite{deap1scint,deap1radon}, have demonstrated the performance characteristics of liquid argon scintillation detectors at the several kilogram scale.  The larger scale dark matter experiments, MiniCLEAN~\cite{miniclean,minicleantaup} and DEAP-3600~\cite{deap3600}, are currently under construction with fiducial masses of 150 kg and 1000 kg, respectively.  Given the exponential energy distribution of nuclear recoils from WIMPs, the ultimate sensitivity of these experiments will depend in part on how effectively \artn events can be rejected at energies of tens of keV or lower.

Argon scintillation is produced by the de-excitation of dimer states created by the ionization track of a particle~\cite{argonscint}.  There is both a short-lived singlet state, $\tau_s$ with a lifetime of $7.0\pm1.0$~ns, and a longer-lived triplet state, $\tau_l$ with a lifetime of $1600\pm100$~ns~\cite{arscinttime}.  The relative amounts of singlet and triplet states produced by a charged particle depend on both the type of recoiling particle and the energy deposited, as measured in~\cite{microclean2008}.  This, along with the large separation of the singlet and triplet time constants, offers excellent particle identification in liquid argon as described in~\cite{boulayhime}.  Using a full simulation of the MiniCLEAN detector (see sections~\ref{sec:deap1sim} and~\ref{sec:sim}) with the above liquid argon scintillation characteristics, we have optimized\footnote{The \fp integral bound optimization minimizes the number of electronic recoils leaking into the $50\%$ nuclear recoil acceptance region in MiniCLEAN Monte-Carlo.} the integration bounds in Equation~\ref{eq:fp}, and we take $T_i=-28$~ns, $\epsilon=80$~ns, and $T_f=15360$~ns in the following sections.

The scintillation light yield of nuclear recoils in liquid argon at energies above 20~keV is $25\pm1$\% of the light yield for electronic recoils~\cite{microclean2012}.  We refer to energies of all types of recoils in units of \emph{electron equivalent} keV (keVee), where this 25\% quenching factor has already been applied to energies for nuclear recoils.  When referring to the full recoil energy, without quenching, we use units of keVr.

\section{Sources of Background Leakage with \fp}
\label{sec:fp_model}

In this section, we highlight the mechanisms in scintillator detectors which contribute to \fp leakage to motivate the potential improvement in a more sophisticated approach.  In order to treat a wide variety of detector contributions to \fp leakage in a consistent manner, we take a predominantly Monte Carlo approach, assuming no functional form for any of the \fp distributions, except at the stage of the initial statistical fluctuations in the number of states produced.

As shown in the following, realistic detector effects can degrade the achievable \fp discrimination between background and signal by many orders of magnitude relative to the fundamental limit, which is set by the statistical fluctuations in the scintillation states produced.  For the case considered here, 12.5~keV apparent energy liquid argon scintillation events in MiniCLEAN, the \fp leakage fraction with 50\% signal acceptance is increased from $3.9\times10^{-11}$ with only fluctuations in the scintillation states produced to $4.2\times10^{-6}$ once all realistic detector effects are included.  In this section, we describe each of the factors that contribute to this lost rejection efficiency.  In the following sections, we develop a new method that regains some of the lost rejection.

For the results in this section, we use a GPU-based fast Monte Carlo simulation to sample and fluctuate the properties of individual photoelectrons from thousands of events in parallel.  This highly parallel approach allows us to reliably sample the tails of the \fp distributions over many orders of magnitude, without having to rely on extrapolation of a phenomenological model.  Detector optics and full simulation of the data acquisition system are bypassed in this simplified model.  In Sections~\ref{sec:results} and~\ref{sec:sim}, we will present results from our full detector simulations of DEAP-1 and MiniCLEAN, that use a more complete model including detector optics and DAQ.

We consider a general two-exponential scintillation model that is equivalent to, but parametrized slightly differently than, that presented in \cite{microclean2008}.  The assumed probability density function for the production times of argon scintillation photons is
\begin{equation}
P(t) = f \left( \frac{1}{\tau_s} e^{-t/\tau_s}\right) + (1 - f) \left( \frac{1}{\tau_l} e^{-t/\tau_l}\right),
\label{eq:scint_time}
\end{equation}
where $f$ is the fraction of states produced with the short-lived time constant\footnote{Although $f$ is sometimes called the ``prompt-fraction,'' it is not equal to the test statistic \fp.}, $\tau_s$, and the remaining states are produced with the longer lived time constant, $\tau_l$.  Generally all three of these parameters depend on the energy and type of the scintillating particle.  For concreteness, we focus on scintillation in liquid argon detectors. However, this model and conclusions drawn from it are applicable to a wide range of scintillators.

For further concreteness in the fast Monte Carlo model, we assume the properties of the MiniCLEAN dark matter detector.  The detector is described in detail in Section~\ref{sec:miniclean}, but its relevant scintillation characteristics are summarized in Table~\ref{tab:detector_param}.  For the purposes of the fast Monte Carlo simulation, the MiniCLEAN detector is considered to be a 500~kg liquid argon target surrounded by 92 optical cassettes.  Each optical cassette houses a Hamamatsu Photonics 8'' R5912-02-MOD PMT and a 10~cm thick acrylic plug.  The acrylic plug acts as a substrate for the TPB layer that converts the extreme UV argon scintillation light~\cite{argonuv1,argonscint} to a wavelength regime where the PMTs are sensitive.

\begin{table}
\begin{center}
\begin{tabular}{|c|c|} \hline
Parameter & Value \\ \hline \hline
Light yield & 6~PE/keVee \\ \hline
Number of photoelectrons & 75~PE \\ \hline
Nuclear quenching factor & 0.25~keVee/keVr \\ \hline
$e^-$ singlet fraction ($f$) & 0.308 \\ \hline
Ar recoil singlet fraction ($f$) & 0.678 \\ \hline
Singlet time const. ($\tau_s$) & 7.0~ns \\ \hline
Triplet time const. ($\tau_l$) & 1600~ns \\ \hline
PMT gain & $3\times 10^7$ \\ \hline
PMT single PE mean charge & 5~pC \\ \hline
PMT timing resolution & 1.5~ns \\ \hline
PMT dark hit rate & 1.5~kHz \\ \hline
Digitizer sample size & 4~ns \\ \hline
Digitizer noise RMS & 0.41~mV \\ \hline
\end{tabular}
\end{center}
\caption{Basic scintillation and detector parameters for 12.5~keVee events in the MiniCLEAN detector.}
\label{tab:detector_param}
\end{table}

Due to the presence of \artn, a beta decay isotope with 565~keVee endpoint, MiniCLEAN will observe $10^{10}$ \artn decays over all energies in a year of running with atmospheric argon, so we focus on this as the primary particle identification background to the signal of interest, WIMP induced nuclear recoils.  For the discussion of the fast Monte Carlo simulation, we will further restrict the discussion to 75~PE (12.5~keVee) apparent energy events in MiniCLEAN which is meant as a feasible but aggressive energy threshold for a WIMP dark matter search.  Refer to Section~\ref{sec:miniclean} for more details on the MiniCLEAN detector.

Leakage is defined to be the fraction of background events with an \fp test statistic greater than the median \fp test statistic for signal.  This yields an approximate 50\% signal acceptance probability.  Except where noted, each leakage fraction is estimated using the very high statistics fast Monte Carlo simulation, including only the detector effects of interest.  This allows trillions of events to be generated in several hours and avoids the need to extrapolate an analytic function to estimate very low leakage levels.  Unless otherwise specified, all leakage fractions have been computed with sufficient statistics to reduce their statistical uncertainty to less than 20\%.  We will use the complete optical simulation to evaluate background leakage in Section~\ref{sec:sim}.

\begin{figure}[htbp]
\begin{center}
\includegraphics[width=1.0\columnwidth]{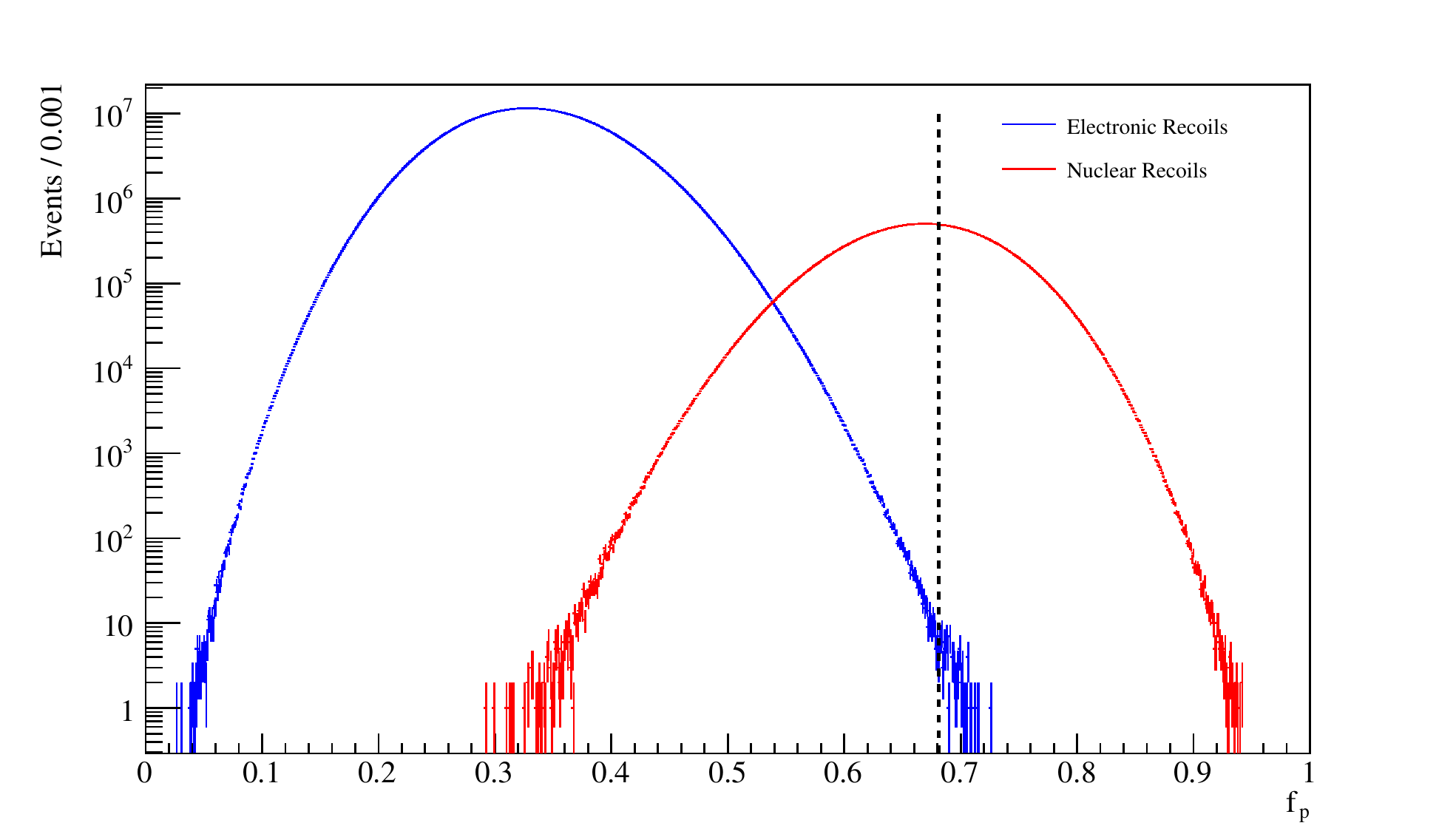}
\caption{\label{fig:gpu_dist} Sample \fp distributions for electronic (blue) and nuclear (red) recoils including the effects up to Section~\ref{sec:pmt_response} (color version online).  The dashed vertical line indicates 50\% nuclear recoil acceptance.}
\end{center}
\end{figure}

In the remainder of this section, we identify various effects that increase the \fp leakage fraction.  The simple estimate of leakage from binomial fluctuations at 75~PE (12.5~keVee) apparent energy would suggest background leakage as low as $3.9 \times 10^{-11}$.  However, including the following realistic detector characteristics, we find that with the simple \fp test statistic one generally can only do as well as $4.2\times 10^{-6}$ at 12.5~keVee with 6~PE/keV detected scintillation light yield.  Samples of the \fp distributions obtained for electronic and nuclear recoils, including only those effects described below up to Section~\ref{sec:pmt_response} are shown in Figure~\ref{fig:gpu_dist}.

\subsection{Binomial Fluctuations}

If the detected photoelectrons could be perfectly identified as coming from the de-excitation of either the singlet or the triplet states, then the only uncertainty in particle identification would come from the Poisson fluctuations in the production and detection of photons from each dimer state.  At a fixed number of photoelectrons, these Poisson fluctuations in the detection of the two states appear as binomial fluctuations in the fraction of photoelectrons coming from the singlet state.  At this stage, with perfect identification of the scintillation states, we estimate \fp using $f$ of Equation~\ref{eq:scint_time}.  With such a simple model, \fp is not a continuous distribution, so 50\% acceptance of nuclear recoils can only be approximated.  The numerically-estimated leakage purely from binomial fluctuations at 75~PE (12.5~keVee) is $3.9 \times 10^{-11}$, with a nuclear recoil acceptance of 46\%.  This represents a fundamental lower bound on the leakage, regardless of analysis technique, and can only be improved by increasing the overall light yield of the experiment.

\subsection{Scintillation time profile}

The originating dimer states that produce each photoelectron cannot be identified perfectly because both states generate photons with overlapping exponential time distributions.  Argon has excellent particle discrimination properties because the time constants are well separated for the two dimer states, but the inherent ambiguity does increase the leakage to $2.9 \times 10^{-10}$.

\subsection{$^{39}$Ar Spectrum and Detector Energy Resolution}
\label{sec:fp_eres}

\begin{figure}
\begin{center}
\includegraphics[width=1.0\columnwidth]{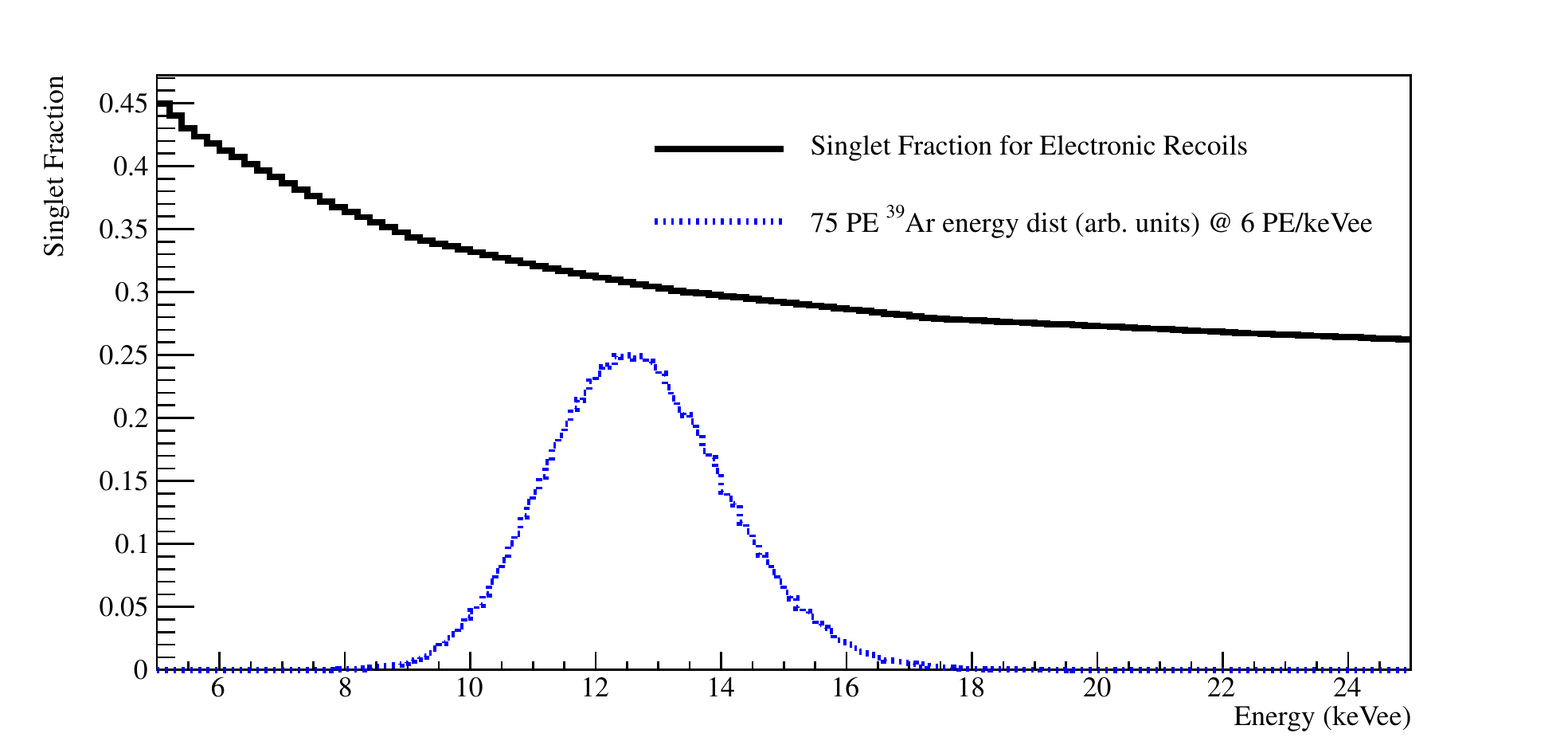}
\caption{Energy dependence of the singlet fraction in electronic recoils~\cite{microclean2008} overlaid with the energy spectrum (in arbitrary units) of 75~PE events from \artn, assuming a detector detection efficiency of 6~PE/keVee.}
\label{fig:singlet_energy}
\end{center}
\end{figure}

The overwhelming source of electron-like backgrounds in a large, well shielded liquid argon detector is the beta decay of \artn.  The finite energy resolution of a real detector allows a distribution of \artn beta energies to be detected with a fixed number of photoelectrons.  Figure~\ref{fig:singlet_energy} shows the true energy distribution for \artn events with an estimated energy of 12.5~keVee in the MiniCLEAN fast Monte Carlo simulation.


The true energy of recoils does not directly impact the statistical fluctuations in the fraction of detected singlet photons, which depends only on the number of photoelectrons.  However, the scintillation properties of argon are energy-dependent, so the energy resolution does directly impact the distribution of singlet photons produced.  Figure~\ref{fig:singlet_energy} shows the energy dependence of the singlet fraction as a function of energy~\cite{microclean2008} with the energy distribution at 75~PE overlaid.  As a result, events with a fixed number of photoelectrons come from a range of expected singlet fractions.  The leakage probability for \artn events with a given number of photoelectrons is dominated by those events whose {\it true} energy is lower than average, corresponding to singlet fractions closer to the 50\% acceptance cut.

This energy-dependence of the scintillation light makes the leakage probability depend on both the number of photoelectrons in the event \emph{and} the light yield of the detector.  Improving the light yield of the detector and holding the energy threshold fixed, thus lowering the energy threshold in keVee, will reduce the background leakage due to the combined effect of higher photoelectron statistics and narrower energy resolution.  Conversely, improving the light yield and holding the energy threshold fixed in number of photoelectrons will slightly increase the background leakage probability due to the energy dependence of the singlet fraction.

For 75~PE events in the MiniCLEAN detector with a light yield of 6 PE/keVee, the finite energy resolution of the detector increases the estimated leakage fraction to $2.3 \times 10^{-9}$.


\subsection{PMT Charge and Time Response}
\label{sec:pmt_response}

\begin{figure}[htbp]
\begin{center}
\includegraphics[width=1.0\columnwidth]{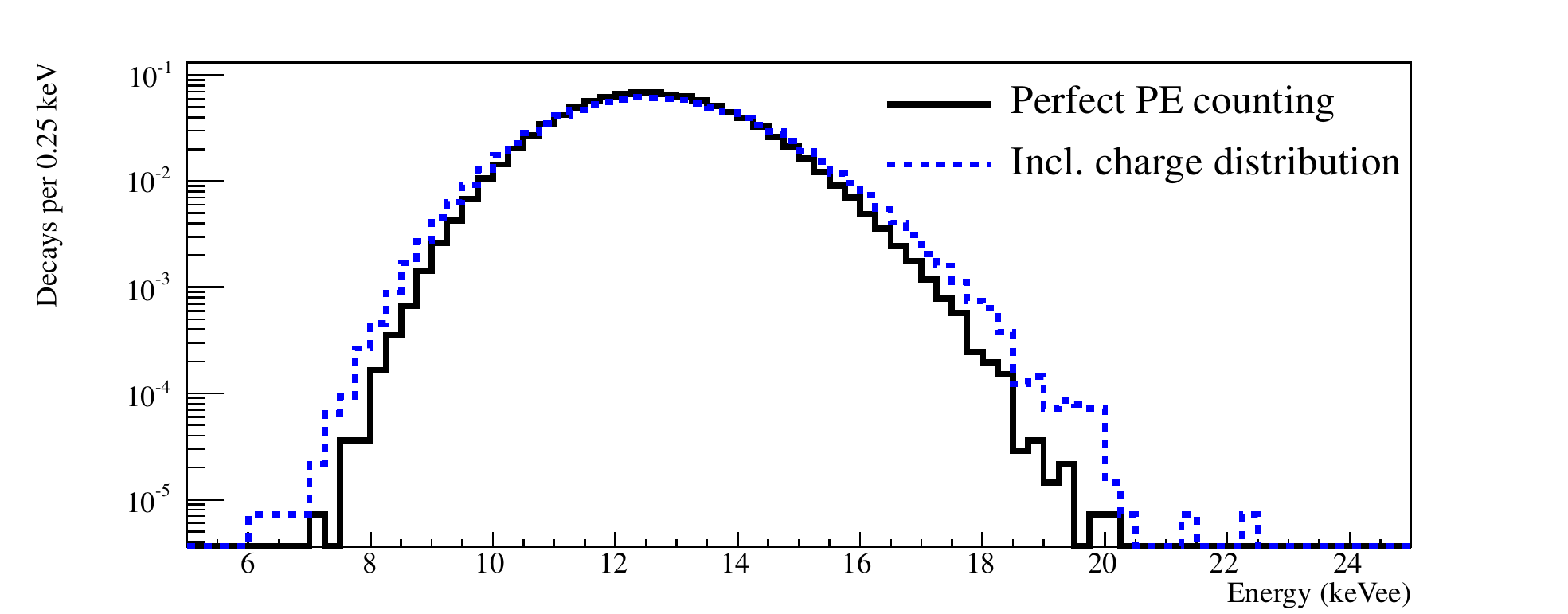}
\caption{\label{fig:eres_charge} Energy distribution of $^{39}$Ar events falling in the 75~PE bin (12.5~keVee) before and after the R5912-02-MOD charge distribution is included.}
\end{center}
\end{figure}

Photomultiplier tubes can produce a range of different pulse sizes given a single photoelectron.  This inherent charge resolution has a two-fold effect on background leakage.  First, it introduces some randomness in the determination of the number of photoelectrons in an event, which must now be defined as the summed charge from all PMTs divided by the PMT gain.  This effectively broadens the energy resolution of the detector, as shown in Figure~\ref{fig:eres_charge}.  The second effect of the charge resolution is to introduce fluctuations in the apparent number of singlet and triplet states beyond the binomial fluctuations inherent in the underlying physics.  Several photoelectrons can individually fluctuate to each have two or three times the mean charge, which broadens the tails of the \fp distribution for both nuclear recoils and electrons.

The time response of a PMT also influences \fp by smearing out the time structure of the scintillation light.  The timing jitter of the R5912-02-MOD is approximately 1.5~ns, but measurements indicate that a photoelectron also has a 4.7\% chance of producing a ``late'' pulse, delayed by up to 60~ns \cite{r591202mod}.  In addition, the finite width of a PMT pulse, as shown in Figure~\ref{fig:pmt_pulse}, makes it possible for the prompt time window in the \fp estimation to only partially contain the charge from the pulse.  The integral in the numerator of Equation~\ref{eq:fp} is relatively insensitive to these effects, although they do contribute in small part to the overall leakage.

To simulate these effects in the fast Monte Carlo simulation, we sampled charges from the full charge distribution, shown in Figure~\ref{fig:eres_charge},  and a simplified Gaussian timing distribution with a RMS of 1.5~ns that ignores the contributions of late pulsing.  For speed reasons, the effect of PMT pulses partially extending beyond the prompt window was estimated by approximating the pulse shape with a Gaussian function.  The \fp leakage of 75~PE electrons increased to $7.1 \times 10^{-8}$.

\subsection{PMT Dark Current}
\label{sec:dark_current}

Thermionic emission of individual electrons inside the PMT from the photocathode and dynode structure produces a ``dark current'' that can affect \fp.   In the MiniCLEAN detector, the 92 PMTs each produce dark pulses at a rate of approximately 1500~Hz at their operating temperature of 87K.  Therefore, dark pulses will be uniformly distributed over the 16~$\mu$s event window, with an average of 2.2 dark pulses per event.  In order to compensate for this offset in the energy scale, we compute the leakage for 77~PE (instead of 75) when dark pulses are included.  The general effect of dark pulses is to further broaden the energy resolution of the detector, which worsens the leakage for the same reasons discussed in Section~\ref{sec:fp_eres}.  In addition, dark hits are most likely to occur outside the prompt time region which biases the \fp of both electronic and nuclear recoils to smaller values, but biases the nuclear recoils more, bringing the two distributions closer together.  As a result, we find that the inclusion of a Poisson distribution of dark hits increases the \artn leakage fraction at 75~PE to $1.2 \times 10^{-7}$.

\subsection{Digitization Noise}

The sampling of the PMT signals introduces an additional source of noise into the calculation of the integrals in \fp.  This noise can come from pickup of analog signals in the PMT cabling to the digitizers or intrinsic noise in the analog-to-digital converters.  The precise impact of digitization noise on \fp depends on the amplitude and frequency composition.  In the fast Monte Carlo simulation, we have introduced this noise as a per-sample Gaussian random variable with a RMS of 0.41 mV, consistent with measurements of the CAEN V1720 digitizers used by MiniCLEAN~\cite{minicleantaup}, DEAP-1~\cite{deap1radon}, and DEAP-3600~\cite{deap3600}.  When zero suppression is enabled in the digitizer readout, the largest contribution of noise to \fp will occur when the number of samples is maximized by separation of photoelectron pulses in time and in different PMT channels.  In the MiniCLEAN detector, with 92 PMTs, this separation of signals in time and PMT channel is quite probable for 75~PE events inside the fiducial volume of the detector.  In this case, each photoelectron pulse is recorded in a separate block of 32 samples, which could contribute an additional Gaussian smearing per photoelectron of up to 3.7\% of the mean single photoelectron charge at 5~pC gain.  We find that this increases the leakage to $4.2 \times 10^{-6}$ in the fast Monte Carlo simulation.

\subsection{Leakage Summary}

\begin{table}
\begin{center}
\begin{tabular}{|c|r|} \hline
Detector Effect & \fp leakage \\ \hline \hline
Binomial Fluctuations & $3.9 \times 10^{-11}$ \\ \hline
Scintillation Time Profile & $2.9 \times 10^{-10}$ \\ \hline
\artn Distrib. + Energy Res. & $2.3 \times 10^{-9}$ \\ \hline
PMT Response & $7.1 \times 10^{-8}$ \\ \hline
PMT Dark Current & $1.2 \times 10^{-7}$ \\ \hline
Digitization Noise & $4.2 \times 10^{-6}$ \\ \hline
\end{tabular}
\caption{\label{tab:leakage}Leakage of \fp for 12.5~keV apparent energy events only in a detector with 6~PE/keV (assuming 50\% nuclear recoil acceptance) as different detector effects are added.  Each row includes all of the effects above it, and the assumptions for each row are described in detail in the text. The optimal value of $\epsilon$, the length of the prompt window in \fp, is between 60 and 70~ns. The uncertainty for all leakages is 20\%.  }
\end{center}
\end{table}

Table~\ref{tab:leakage} shows the increasing leakage from an \fp cut as more detector effects are included in the fast Monte Carlo model.  Clearly, energy resolution is a significant factor in the leakage due to the increased singlet fraction at lower event energies.  The detector energy resolution comes from a mixture of photon counting statistics, which can only be improved by increasing the light yield of the detector, and the PMT and waveform digitizer response, which can be improved with better waveform analysis techniques, as shown in the next section.

\section{Bayesian Photoelectron Counting}
\label{sec:spe_counting}

Several of the detector effects described in the previous section that increase the \fp leakage fraction can be mitigated with an analysis that can extract individual photoelectron times from the PMT waveforms.  However, rather than solve a general deconvolution problem, we have instead developed an analysis that assumes the PMT waveform is produced by argon scintillation from a single recoiling particle.  This prior can be incorporated into Bayes' Theorem in order to estimate the number of photoelectrons in the waveform, and then assign times to each of the photoelectrons based on the waveform shape.

The event analysis algorithm has 5 stages:
\begin{enumerate}
\item Time/voltage calibration of waveforms.
\item Identification of regions that contain pulses.
\item Bayesian identification of photoelectrons.
\item Reconstruction of the event position and energy.
\item Repeat of Bayesian identification of photoelectrons with improved priors.
\end{enumerate}
These steps are explained in each of the following subsections.

\subsection{Waveform Calibration}
\label{sec:wave_cal}

With the PMT waveforms in a zero-suppressed voltage-time series, the absolute time offset of the waveforms in the event are adjusted so that the summed waveform reaches a maximum amplitude at $t=0$.  Due to the shortness of the singlet time constant, aligning based on the peak removes most of the jitter caused by the latching of the trigger, which can often be several digitizer samples or more.

The voltage of each PMT waveform is separately corrected to remove any baseline offset.  For this first stage of analysis, the first 4~pre-samples from each block of samples in a given channel are averaged to estimate a constant baseline that is subtracted from all the blocks in the channel. After voltage calibration, \fp can also be calculated for the summed PMT waveform, as it is used in the Bayesian photoelectron identification stage.

\subsection{Pulse Finding}
\label{sec:pulse_finding}
\begin{figure}
\begin{center}
\includegraphics[width=1.0\columnwidth]{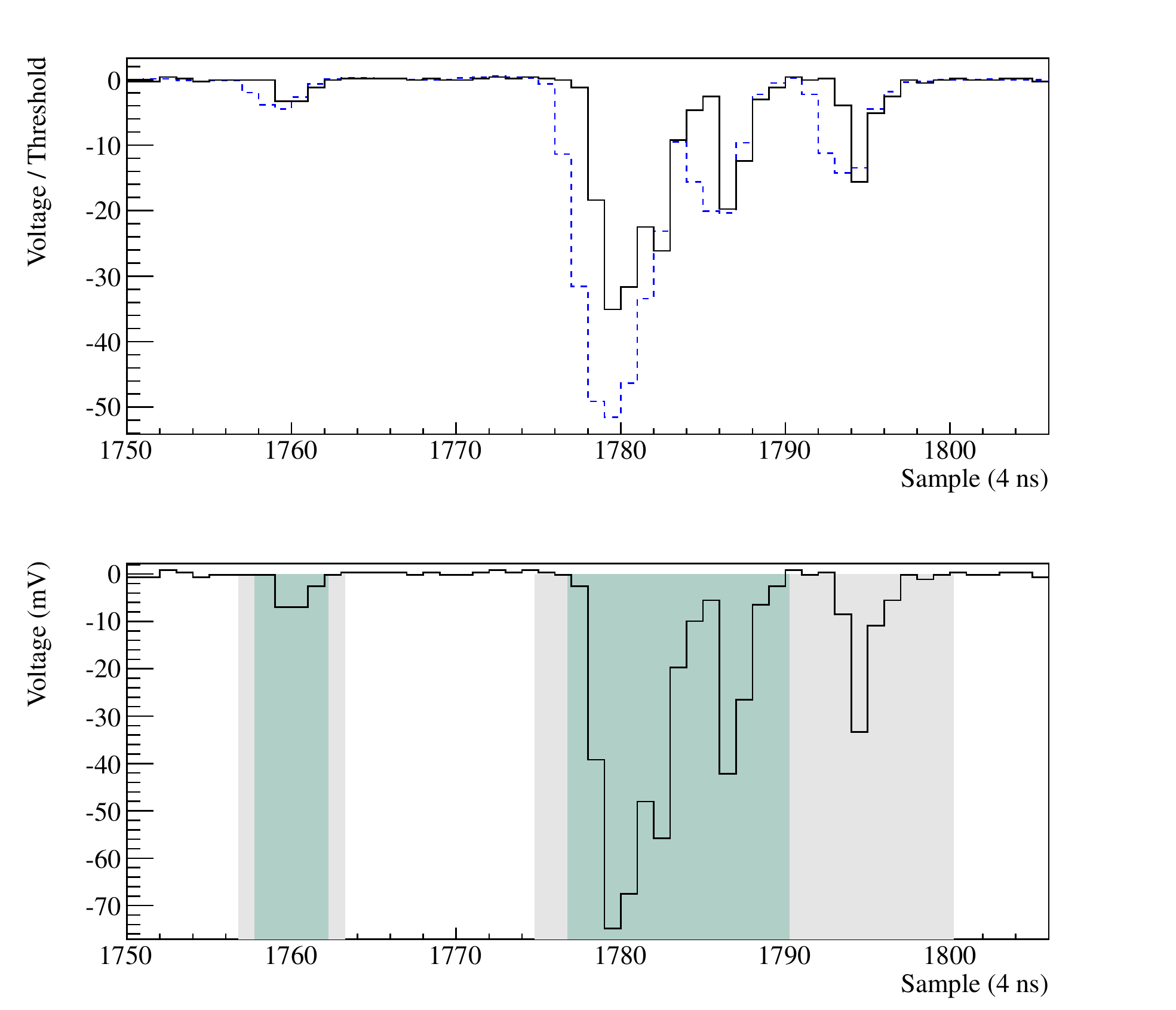}
\caption{\label{fig:sliding_window}A typical voltage waveform from a single PMT in MiniCLEAN Monte Carlo simulation.  The top panel shows the waveform normalized by 5 times the RMS of the electronics noise profile (black, solid) compared to the sliding integral value normalized by the corresponding threshold (blue, dashed).  The sliding integration window enhances the right-skew PMT pulses relative to threshold while providing a filter for high frequency electronics noise.  The bottom panel shows the pulse regions identified by the pulse finder.  Green shaded regions are the regions where threshold is crossed, and the gray regions indicate a buffer region that extends the pulse boundaries.  If threshold is crossed again within the buffer, the pulse boundary is further extended as in the right most pulse region.}
\end{center}
\end{figure}

Single photoelectron pulses from the 8" R5912-02-MOD PMTs typically span 20 ns, as shown in Figure~\ref{fig:pmt_pulse}.  We scan the calibrated waveforms for each PMT separately with a sliding 12~ns (3 sample) integration window and extract a \emph{pulse region} whenever the integral exceeds 5 times the RMS of noise samples times the square root of the number of samples in the window.  Once that detection threshold has been crossed, the boundaries of the pulse region are the times where the sliding window integral drops to below the RMS divided by the square root of the number of samples.  Figure~\ref{fig:sliding_window} shows the sliding window calculation applied to a pulse from 5 photoelectrons.

We find that the sliding window integration method is much more sensitive to low charge PMT pulses without significantly increasing the number of false positive detections.  The width and right skew of the pulses make it difficult to efficiently discriminate low charge pulses from background noise with a simple voltage threshold.  An integration window, however, effectively filters out the high frequency noise while retaining the relatively low frequency PMT signal.

Once the samples corresponding to pulses have been identified, they are removed from the waveform, and the remaining baseline-only samples are used to estimate a local baseline for each block of samples in the original waveform.  Unlike in the calibration stage, the baseline computed in this stage is allowed to vary with time, thus removing any lower frequency components, for example under- or over-shoot in PMT base electronics.  The new baseline is subtracted from the waveform and the entire pulse finding procedure is repeated.  After two passes, the results are stable and robust against low-frequency baseline variations.

\subsection{Bayesian Photoelectron Identification: First Pass}
\label{sec:bayes_1st_pass}

Once pulse regions have been extracted from each PMT waveform, the next task is to identify how many separate photoelectron pulses are contained in each pulse region, and what their arrival times were.  As discussed in Section~\ref{sec:intro}, two natural techniques to apply are normalized integral charge, where the integral of each pulse region is divided by the average single photoelectron charge, or peak counting, where the number of local maxima in the pulse region are counted after some filtering.  Charge integration is a good strategy to apply at early times, where pileup will likely make multiple peaks indistinguishable, and a poor strategy at late times, where a pulse region will almost always contain a single photoelectron, but the variance in the single photoelectron charge distribution will create large fluctuations in the estimate.  Conversely, peak counting is a poor strategy at early times, where it will generally undercount photoelectrons, but a good strategy at late times, where it is insensitive to charge fluctuations in single photoelectrons.

Bayes' Theorem provides a quantitative way to incorporate our knowledge of the time structure of argon scintillation light.  For each pulse identified in each PMT waveform we would like to know the most probable number of photoelectrons in the pulse, $n$, given that the pulse spans time $t_1$ to $t_2$ (where $t_1$ and $t_2$ are defined by the pulse finding procedure in Section~\ref{sec:pulse_finding}) and has an integrated charge of $q$.  The probability mass function for $n$ is
\begin{align}
\label{eq:p_n_q}
P_N(n|q,t_1,t_2) & =  \frac{P_Q(q \,|\, n) P_N(n \,|\, t_1,t_2)}{P_Q(q \,|\, t_1,t_2)} \nonumber \\
 & =  \frac{P_Q(q \,|\, n) P_N(n \,|\, t_1,t_2)}{\sum_{i=0}^\infty P_Q(q \,|\, i)P_N(i \,|\, t_1,t_2)},
\end{align}
where $P_N$ is used to denote a probability mass function for the number of photoelectrons, and $P_Q$ denotes a probability density function for integrated charge.

The function $P_Q(q \,|\, n)$ is the $n$-photoelectron charge distribution for the PMT, which is assumed to have no dependence on the time of the pulse.  The charge distribution of noise, $P_Q( q \,|\, n=0)$, can be found by applying the pulse finding algorithm to a sample of waveforms with electronics noise only.  The single photoelectron charge distribution, $P_Q(q \,|\, n=1)$, can be measured for each PMT using a triggered, low intensity light source, as in Figure~\ref{fig:charge_dist}.  For $n > 1$, $P_Q(q \,|\, n)$ is the convolution of $P_Q(q \,|\, n=1)$ with itself $n$ times.

\begin{figure}
\begin{center}
\includegraphics[width=1.0\columnwidth]{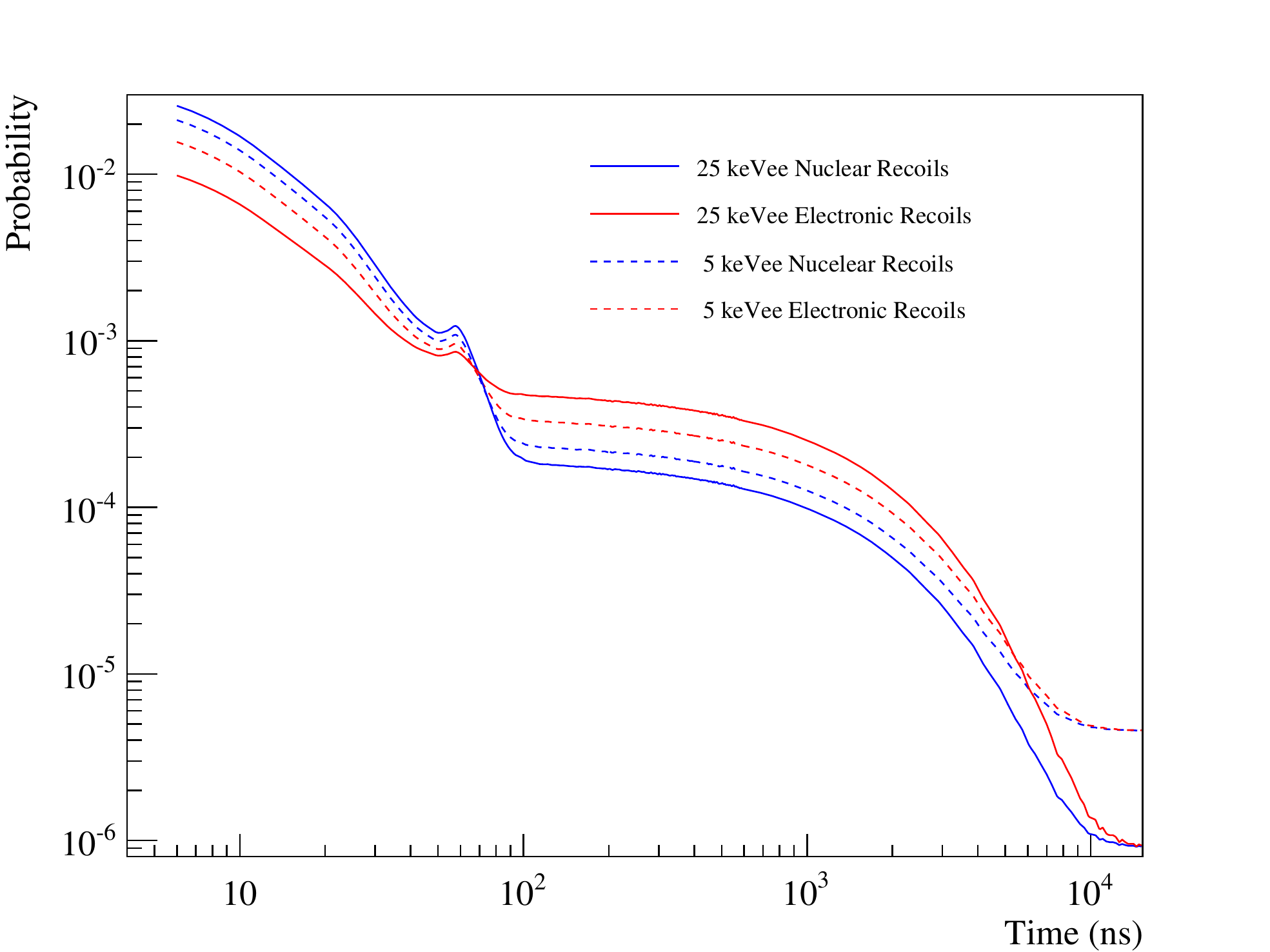}
\caption{\label{fig:time_pdf}Photoelectron detection time PDFs for electronic and nuclear recoils at 5~keVee and 25~keVee observed energies from MiniCLEAN Monte-Carlo simulation (described in detail in Section~\ref{sec:miniclean}).  The peak near 60~ns is due to PMT double and late pulsing~\cite{r591202mod}.  The energy dependence of the mean triplet fraction~\cite{microclean2008} is included in construction of the PDFs.  The flat component in time, due to PMT dark hits, contributes a larger fraction to the signal at low energies resulting in convergence to a higher probability at late times.  In the Bayesian photoelectron counting procedure, the PDFs are constructed by linear combination of the singlet and triplet components with a bootstrapped prior to using the \fp test statistic.}
\end{center}
\end{figure}

Our knowledge of the distribution of scintillation photons (and therefore the probability of pileup) enters into the equation via $P_N(n \,|\, t_1,t_2)$, the probability mass distribution for the number of photons in the given pulse.  Suppose that events of a particular class produce $\mu$ detected photoelectrons in the PMT on average with a photoelectron detection time probability density function (PDF) of $S(t)$.  The probability mass function of observing $n$ photoelectrons in the time interval $[t_1, t_2]$ is then
\begin{equation}
\label{eq:p_n}
P_N(n \,|\, t_1,t_2) = \sum_{j=0}^{\infty} \mathrm{Pois}(j \,|\, \mu) \times \mathrm{Bin}(n \,|\, j, I_j),
\end{equation}
where $\mathrm{Pois}(j \,|\, \mu)$ is the Poisson probability of observing $j$ photoelectrons in the entire waveform given the expected value $\mu$, $\mathrm{Bin}(n \,|\, j, I_j)$ is the binomial probability of detecting $n$ photoelectrons given $j$ photoelectrons in the waveform, and the probability, $I_j$, of $j$ photoelectrons falling in the time interval $\left[t_1, t_2\right]$ is defined by
\begin{equation}
\label{eq:I}
I_j = \int_{t_{j,1}}^{t_{j,2}}S(t)dt.
\end{equation}
The PMT timing response (including dark hits, double pulsing, late pulsing, etc~\cite{r591202mod}) and the effects of detector optics are convolved with the scintillation time structure, Equation~\ref{eq:scint_time}, to include realistic detector effects in the photoelectron detection time PDF, $S(t)$.  Figure~\ref{fig:time_pdf} shows sample time PDFs for nuclear and electronic recoils at energies of 5 and 25~keVee in the MiniCLEAN Monte Carlo simulation (described in detail in Section~\ref{sec:miniclean}).  

The integration bounds of Equation~\ref{eq:I}, $t_{j,1}$ and $t_{j,2}$, depend on the hypothesized number of photoelectrons, $j$, due to finite width of the PMT pulses.  For the hypothesis of a single photoelectron spanning the pulse bounds, $t_{1}$ to $t_2$, the known pulse shape can be used to narrow the integration bounds in Equation~\ref{eq:I} to the time interval in which the scintillation photon may have been produced, $\left[t_{1,1}, t_{1,2}\right]$.  However, in the limit of many photoelectrons spanning the same time interval, the scintillation photons may have been produced over the entire range of the pulse, $\left[t_1, t_2\right]$.  In practice, we compute the cumulative distribution function of the time PDF, $S(t)$, and determine the $1^{\mathrm{st}}$ and $j^{\mathrm{th}}$ $(j+1)$-quantiles, $q_1$ and $q_j$ respectively.  Letting $\tau$ represent the finite width of the digitization sampling, we then take $t_{j,1}=q_1-\tau/2$ and $t_{j,2}=q_j+\tau/2$.  This gives a robust estimation of the correct integration boundaries for $j$ photoelectrons without requiring exact knowledge of the PMT pulse shape with $j$ photoelectrons piling up with random time offsets.  By inserting Equation~\ref{eq:I} into Equation~\ref{eq:p_n} and Equation~\ref{eq:p_n} into Equation~\ref{eq:p_n_q}, $P_N(n|q,t_1,t_2)$ can be evaluated.

The selection of $\mu$ and $S(t)$ for each PMT (see Figure~\ref{fig:time_pdf}) depends on our hypothesis of the particle type, position and energy, but estimating those quantities in turn requires the results of photoelectron counting.  In order to bootstrap the process, we perform this first pass of the analysis assuming that $\mu$ for each PMT is equal to the total integrated charge observed by the PMT divided by the average single photoelectron charge.  The time distribution for photoelectrons is estimated by using the prompt-fraction test statistic, \fp, which is computed using all the PMT waveforms.  Separate time PDFs for the singlet and triplet photons generated with the Monte Carlo simulation are linearly combined to create $S(t)$ according to the singlet fraction calculated from \fp.  A flat time distribution for PMT dark hits is then added into $S(t)$ based on the relative magnitudes of the expected number of PMT dark hits and $\mu$.  As described in Section~\ref{sec:wave_cal}, $t=0$ for each event is defined as the peak in the waveform summed over all PMTs in order to be consistent with the definition of $S(0)$ as the maximum of the time PDF.

\begin{figure*}[t]
\centering
\begin{subfigure}[t]{0.45\textwidth}
\includegraphics[width=\textwidth]{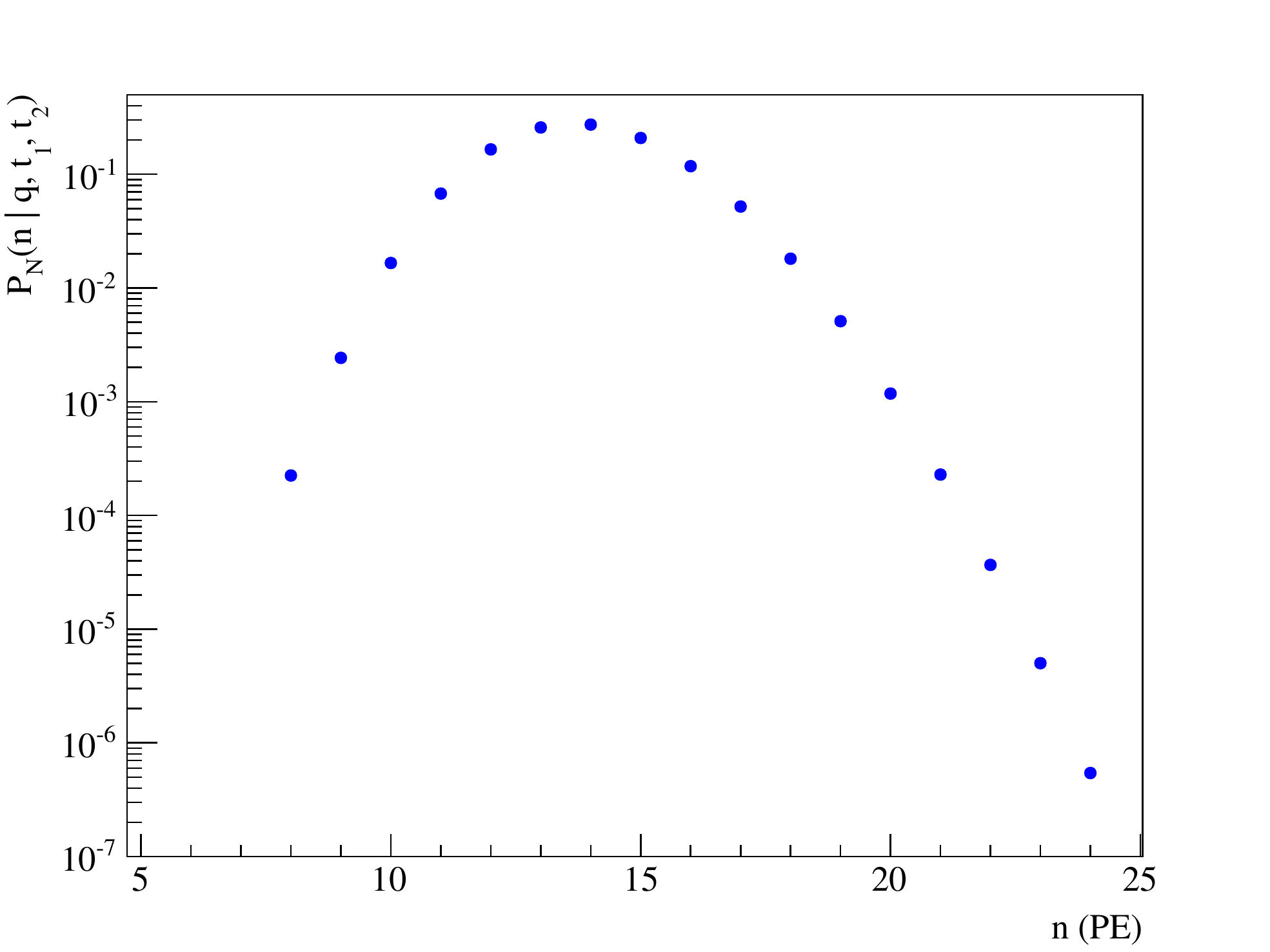}
\end{subfigure} \qquad
\begin{subfigure}[t]{0.45\textwidth}
\includegraphics[width=\textwidth]{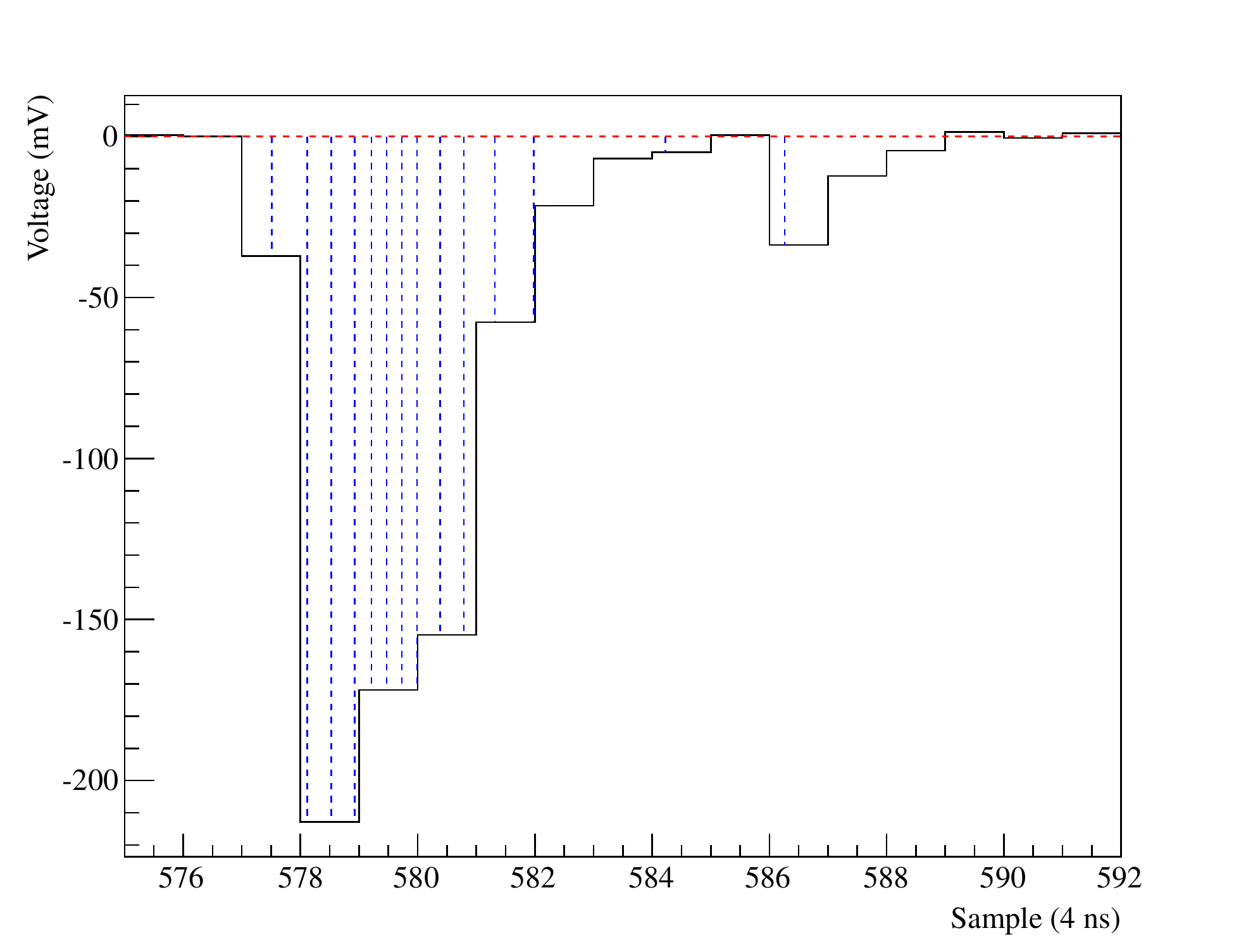}
\end{subfigure}
\caption{\label{fig:quantiles}(Left) Sample distribution of $P_N(n | q,t_1,t_2)$ for the pulse shown in the right panel.  The Bayesian photoelectron counting algorithm assigns 14 photoelectrons to this pulse.  (Right) The assigned times using the waveform shape are shown by the vertical blue dashed lines.}
\end{figure*}

With these bootstrap priors, the most probable number of photoelectrons in each pulse region can be estimated as the integer $n$ which has maximal $P_N(n | q,t_1,t_2)$ from Equation~\ref{eq:p_n_q}.  In order to assign times to the photoelectrons, the cumulative distribution function  (CDF) of the waveform in the region is used.  For a pulse region with most probable $n$ photoelectrons, we determine the $n+1$-quantiles of the model CDF.  Each photoelectron is assigned a unique detection time equal to one of these quantiles.  Interpolation is used so that the quantiles can be between samples.  This procedure is demonstrated in Figure~\ref{fig:quantiles} for a pulse region with 14 photoelectrons.

\subsection{Position and Energy Reconstruction}

Given counts and times for photoelectrons detected by every PMT in the detector, more event properties can be reconstructed.  With a radius of 43.7~cm, the MiniCLEAN detector is small enough that time is a weak handle on event position, so only the number of photoelectrons observed by each PMT is used by a maximum likelihood algorithm to estimate the event position and energy.  The likelihood function contains a fast, Monte Carlo-derived optical model of the detector (which approximates the detailed optics of the optical cassettes) that can predict the expected number of photoelectrons for each PMT given a hypothesized position and energy.  Once the fit has found the most likely event position and energy, a final set of expected numbers of photoelectron are calculated for each PMT and passed to the next stage of the analysis.

\subsection{Bayesian Photoelectron Identification: Second Pass}

For second pass of the Bayesian photoelectron calculation, we repeat the calculation described in Section~\ref{sec:bayes_1st_pass}, but with an estimate of $\mu$ for each PMT derived from the position and energy reconstruction stage.  The estimate of the time PDF, $S(t)$, is left unchanged, except for the PMT dark hit contribution\footnote{In a larger detector than MiniCLEAN, a different $S(t)$ could be computed for each PMT that includes the distribution of photon propagation times for an event at the reconstructed position.}.  By replacing the crude Bayesian prior from the first pass with a much more accurate one based on event reconstruction, the second pass of the algorithm can reduce the energy bias and improve the energy resolution.  In principle, additional iterations of position/energy reconstruction and Bayesian photoelectron counting could be performed, but we generally find no significant improvement after 2 passes.

\section{Test Statistics for Particle Identification}
\label{sec:stat_tests}

With a set of detected photoelectron times for an event, $\mathcal{T}$, we can compute a variety of different test statistics.  The simplest test statistic is a discrete version of \fp,
\begin{equation}
r_p = \frac{ | \{t \,|\, t \in \mathcal{T} \;\wedge\; T_i < t < \epsilon \} | } { | \{t \,|\, t \in \mathcal{T} \; \wedge\; T_i < t < T_f\} |},
\end{equation}
where $\epsilon$ sets the prompt window, just as in the definition of \fp.  The \rp test statistic removes some of the variance induced by the PMT charge distribution on \fp, but does not take full advantage of the scintillation time structure.

A more powerful approach to separating electronic and nuclear recoils is a likelihood ratio test statistic, as was also briefly described in \cite{microclean2008}.  By evaluating the likelihood of the observed photon detection times using time PDFs for a nuclear recoil hypothesis and an electronic recoil hypothesis, one can make better use of the full timing information from each photon.   We define this second test statistic, \lr, as a normalized log-likelihood difference, comparing the nuclear recoil hypothesis to the electronic recoil hypothesis.  Specifically, we define the test statistic to be:
\begin{equation}
l_r = \frac{1}{m} \sum_{t \in \mathcal{T}}\left( \log P_n(t \,|\, E) - \log P_e(t \,|\, E) \right),
\end{equation}
where $m = |\mathcal{T}|$, $P_n$ is the time PDF for the nuclear recoil hypothesis given an event energy $E$, and $P_e$ is the time PDF for the electronic recoil hypothesis given an event energy $E$.  Sample distributions for $P_n$ and $P_e$ are shown in Figure~\ref{fig:time_pdf} at 5 and 25~keVee.  Due to the sign convention adopted in the definition of \lr, positive values are more nuclear recoil-like and negative values are more electronic recoil-like.  The division by $m$ is a convenience to keep the range of \lr similar for events with different numbers of photoelectrons.  The time PDFs $P_n$ and $P_e$ are computed using separate Monte Carlo simulations of singlet and triplet scintillation photons, which are then linearly combined according to the measured energy (including the quenching factor where appropriate) and particle dependence of the singlet fraction.

Finally, it is always important to pair a likelihood ratio with a goodness-of-fit metric to reject events that do not conform to either hypothesis.  Given the time PDFs $P_n$ and $P_e$ used to compute \lr, we can compute a Kolmogorov-Smirnov test statistic for the nuclear recoil ($K_n$) and electron recoil ($K_e$) hypotheses, respectively.

\section{Bayesian Photoelectron Counting in DEAP-1}
\label{sec:results}

\begin{figure}
\begin{center}
\includegraphics[width=1.0\columnwidth]{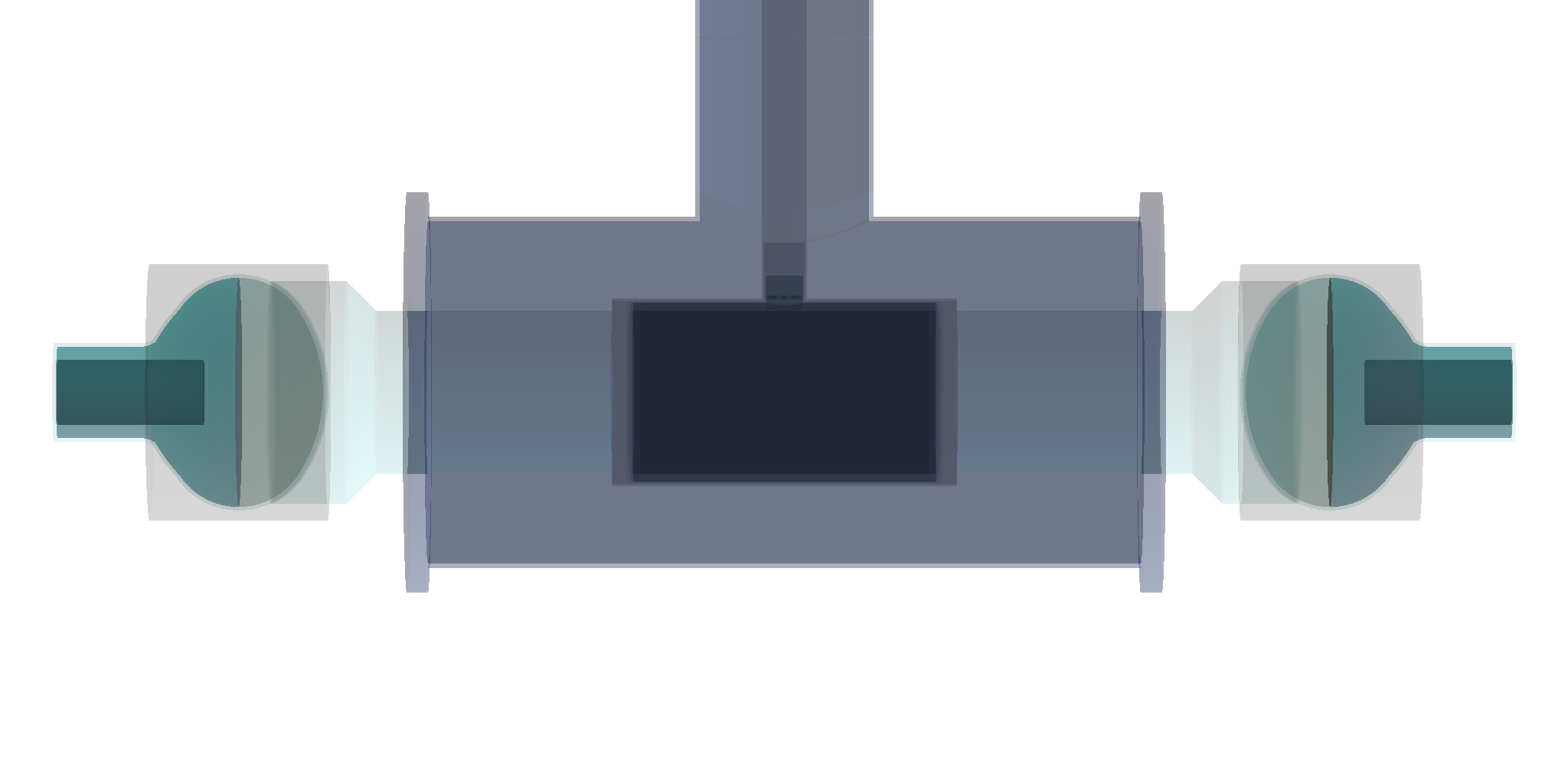}
\caption{\label{fig:deap1}A rendering of the DEAP-1 detector from the GEANT4 simulation.  The 7.6 kg liquid argon volume is coupled to acrylic light guides via TPB coated glass windows.  Each light guide is coupled through a layer of mineral oil to Hamamatsu Photonics R5912 PMT. }
\end{center}
\end{figure}

To demonstrate the effectiveness of the Bayesian photoelectron counting technique, we apply it here to data collected by the DEAP-1 detector between November 28 and December 9, 2011.  DEAP-1 is a cylindrical detector containing 7.6 kg of liquid argon that has been operated underground at SNOLAB since 2007.  The detector has had various configurations as an R\&D platform as described in~\cite{deap1radon}, but the configuration in November-December 2011 is shown in Figure~\ref{fig:deap1}.  

The central target volume is defined by a 28~cm long and 15~cm diameter acrylic cylinder that has been coated on the inside with TPB which converts 128~nm extreme UV light~\cite{argonuv1,argonscint} into 440~nm visible light~\cite{tpbreemit} that can be transmitted through acrylic light guides which are coupled to PMTs through a layer of mineral oil.  The TPB thickness on each acrylic end cap is 0.9~$\mu$m and is coupled via a Kodial glass window and acrylic light guide to a high quantum efficiency (HQE) model R5912 8" PMT manufactured by Hamamatsu Photonics.  The TPB coating on the inside of the acrylic barrel is 4-5~$\mu$m thick, and the outside of the barrel is wrapped with a PTFE reflector.  

In this configuration, DEAP-1 digitizes full 16~$\mu$s waveforms with no zero-suppression using a CAEN V1720 250 MHz digitizer.  Zero-suppression is applied in software to be consistent with the zero-length encoding assumed in the fast Monte Carlo simulation in Section~\ref{sec:fp_model} and the full MiniCLEAN simulation in Section~\ref{sec:sim}.

\subsection{DEAP-1 \na Calibration Data}

The DEAP-1 data set used for this analysis comes from a tagged, 10 $\mu$Ci \na radioactive source placed just outside the acrylic vacuum vessel, centered along the axis of the target volume.  The dominant \na decay mode produces a positron that annihilates into back-to-back 511~keV gamma rays followed within ps by a 1274~keV de-exitation gamma ray.  A small NaI crystal is placed behind the source to tag one of the 511~keV gamma rays in order to constrain the other 511~keV gamma to be within a 9$^{\circ}$ cone aimed directly at the center of the argon volume.  The \na source geometry and trigger configuration used in this data is very similar to that described in \cite{deap1scint}, but the trigger only requires a coincidence between the two main PMTs facing the argon and the small PMT attached to the back NaI crystal.  The 1274~keV gamma ray is untagged and uncorrelated in direction with the 511~keV gamma ray that enters the DEAP-1 target volume.


The light yield of the detector in this configuration was measured to be 4.5~PE/keV using the full energy peak from the 511~keV calibration gammas, and this is incorporated into the Bayesian counting procedure to include the energy dependence of the mean singlet fraction in liquid argon scintillation.  From test bench measurements, for approximately 5\% of photoelectrons, the HQE R5912 PMTs in DEAP-1 produce ionic after-pulses over timescales of several~$\mu$s after the primary pulse.  The time PDFs used in the Bayesian photoelectron counting, and in the calculation of \lr, include these effects with PDFs generated in the simulation described in the next section.  With only 2 PMTs (rather than 92 as in the fast Monte Carlo simulation of MiniCLEAN in Section~\ref{sec:fp_model}), there is significant pileup of photoelectron pulses which provides a good test of the Bayesian photoelectron counting transitioning from pileup of multiple photoelectrons in the prompt region to isolated pulses at late times as described in Section~\ref{sec:bayes_1st_pass}.

DEAP-1 is not capable of 3D position reconstruction, but can reconstruct the position ($z$) of the event along the axis of the target cylinder.  The $z$ position is also used to cut background that occurs at the windows where the PMT light guides are coupled to the liquid argon target volume (see~\cite{deap1radon} for details).  Only events reconstructing within 10~cm of the $z$ center of the detector are included in the analysis.  As a data quality cut and to remove pileup of multiple recoils in the detector, we also apply a cut using the KS test statistics $K_e$ and $K_n$ described in Section~\ref{sec:stat_tests}.

\begin{figure*}
\begin{center}
\includegraphics[width=1.0\textwidth]{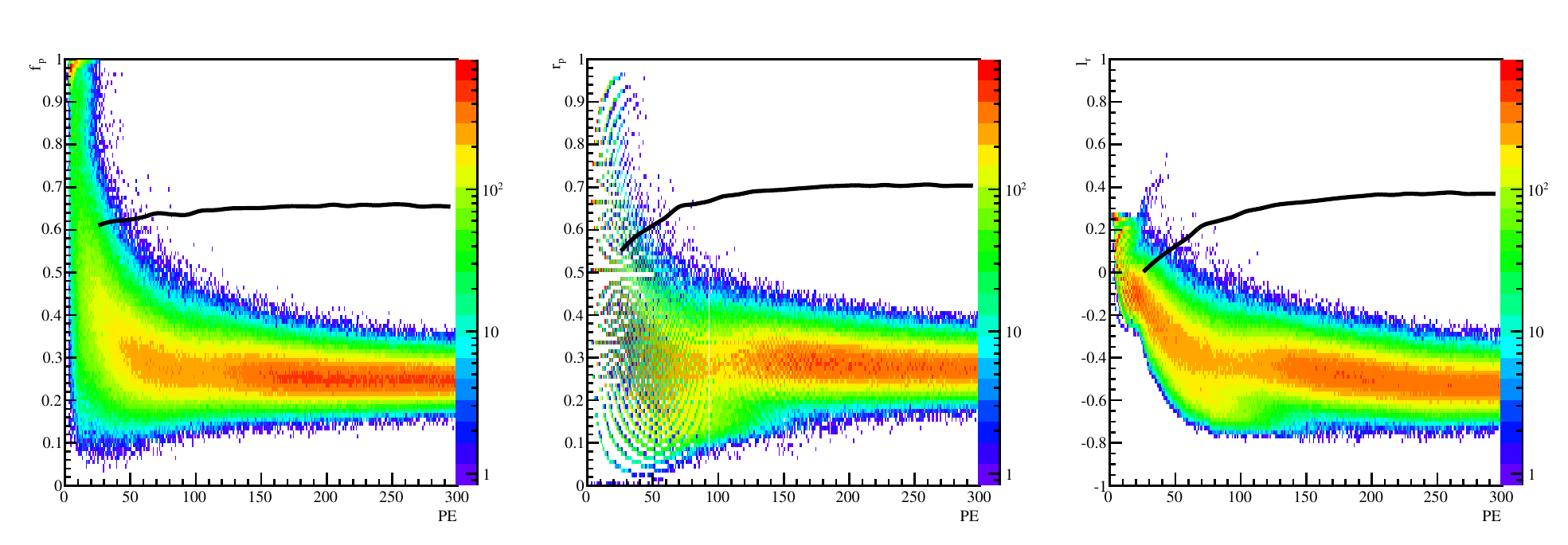}
\includegraphics[width=1.0\textwidth]{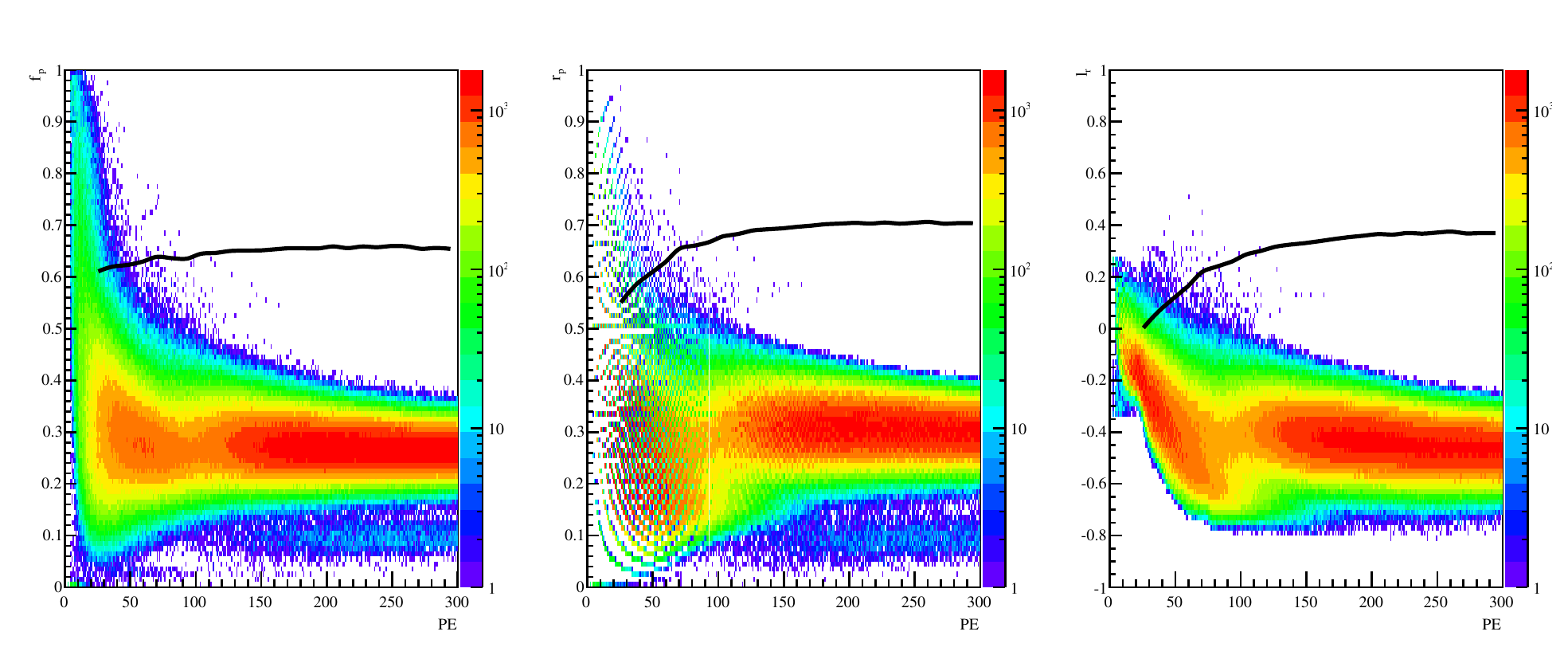}
\caption{\label{fig:deap1_2d}Distribution of \fp (left column), \rp (middle column), and \lr (right column) test statistics for simulated electronic recoils from a \na calibration source as a function of reconstructed number of photoelectrons in DEAP-1 simulation (top row) and data (bottom row).  The solid black line in each panel represents the simulated mean profile of each test statistic for nuclear recoils in DEAP-1.  The discrete nature of the \rp test statistic creates the structure which is apparent in the center panels.  A population of events at positive \lr values below 100 PE due to pileup of Cherenkov light from the 1274~keV gamma with scintillation from the 511~keV gamma becomes increasingly apparent with the improved test statistic.}
\end{center}
\end{figure*}

The bottom panels of Figure~\ref{fig:deap1_2d} show the distributions of the three test statistics, \fp, \rp, and \lr, as a function of the number of detected photoelectrons as output by the Bayesian photoelectron counting procedure.  Events at low number of photoelectrons and high values of \fp which could leak into the nuclear recoil region are due in part to the effects described in Section~\ref{sec:fp_model}.  However, an additional class of high \fp events are also present in the region of 25 to 150 PE.  The presence of this class of events is enhanced with the Bayesian photoelectron counting procedure and the \rp test statistic.  Using the \lr statistic, which takes full advantage of the scintillation time profiles, these events are especially well identified at positive values of \lr.  As described in the next section, simulations suggest that these events are due to the untagged 1274~keV gamma from the \na decays producing Cherenkov light in the large acrylic light guides in coincidence with the 511~keV gamma producing scintillation light in the liquid argon.  The fast Cherenkov light produces an additional prompt component which the \fp test statistic is relatively insensitive to.  The separation of these events from the scintillation only events highlights the improved particle identification capabilities of the Bayesian photoelectron counting technique and the \lr test statistic.

\begin{figure}
\begin{center}
\includegraphics[width=1.0\columnwidth]{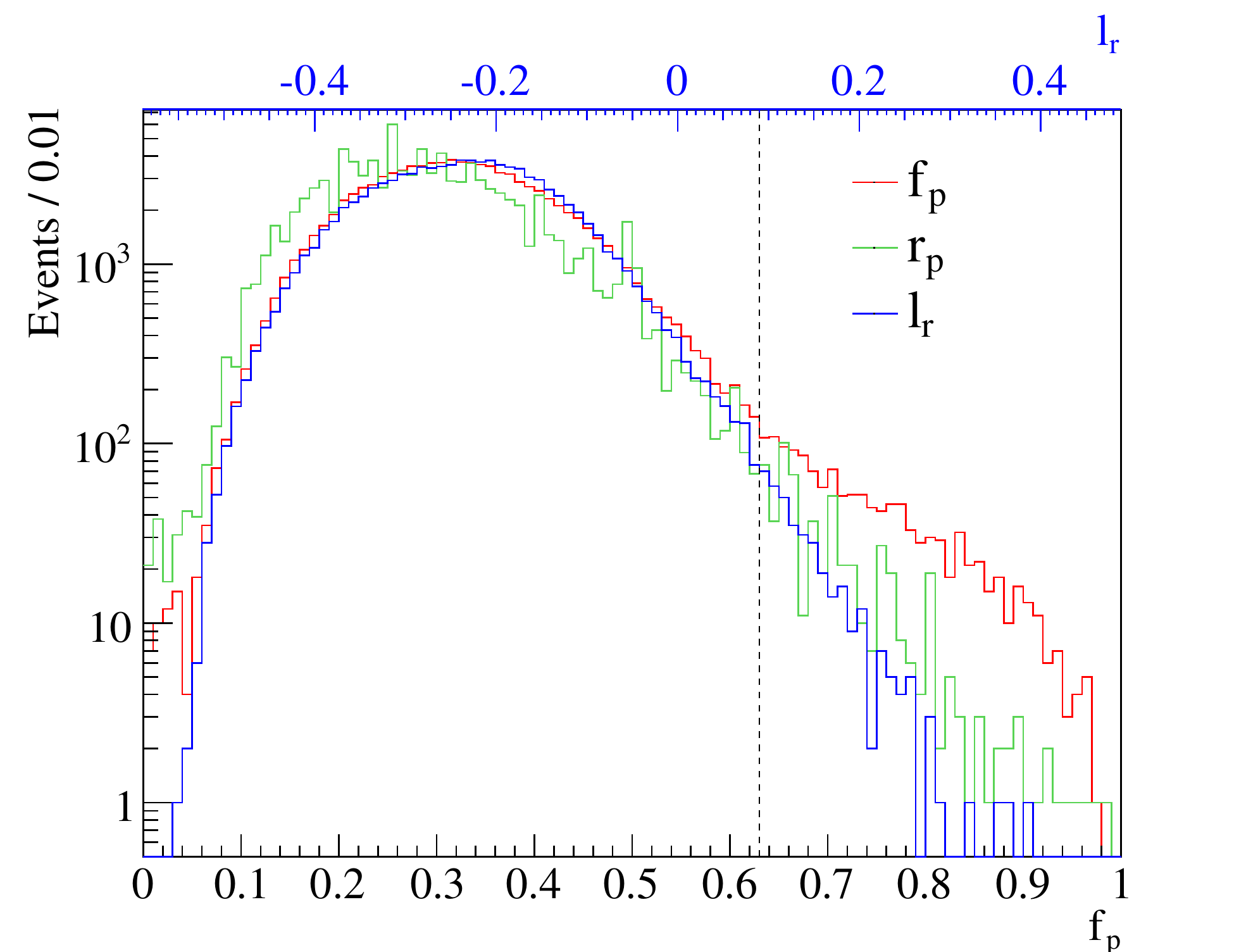}
\caption{\label{fig:deap1_1d_compare}Distribution of \fp, \rp, and \lr test statistics for electronic recoils for \na calibration events in DEAP-1 with 30 PE.  The vertical dashed line indicates 50\% nuclear recoil acceptance at 6.7~keVee.  The \lr values have been linearly transformed such that the median values for the electron and nuclear recoil distributions match those for \fp.  The shift in the \rp peak relative to \fp is due to the discrete nature of the test statistic.}
\end{center}
\end{figure}

A one-dimensional slice of the distributions at 30~PE, or 6.67~PE/keVee, is shown in Figure~\ref{fig:deap1_1d_compare}.  For the purposes of comparison, the \lr values have been linearly transformed to match the median values for electronic and nuclear recoils for \fp.  This keeps a fixed 50\% nuclear recoil acceptance for each test statistic.  The reduction in the \lr tail relative to \fp on the nuclear recoil side shows that \lr is a superior background rejection tool for low energy events.  Integrating above \fp of 0.7, the Bayesian photoelectron counting and the \lr test statistic reduce the leakage of 30~PE events into the nuclear recoil 50\% acceptance region by a factor of $7.8\pm0.3\;\mathrm{(stat)}$.

As in~\cite{deap1scint}, a well-motivated model can be constructed to extrapolate the tails of the \fp distribution at a particular energy.  However, we lack such a model for the normalized likelihood ratio discriminant \lr.  For this reason, we avoid comparison of the leakage for the particle identification parameters \fp and \lr except where there is sufficient statistics in the data to make a direct comparison.

\subsection{DEAP-1 Simulation}
\label{sec:deap1sim}

To further test the Bayesian photoelectron counting technique and the test statistics \fp, \rp, and \lr, we apply the complete procedure described in Section~\ref{sec:spe_counting} to simulated DEAP-1 \na calibration data in this section and to \artn events in MiniCLEAN in Section~\ref{sec:sim}.  The simulation and analysis framework used for both analyses, called \emph{RAT}, was originally developed by the Braidwood collaboration~\cite{braidwood} and is now maintained cooperatively by the DEAP/CLEAN collaborations and the SNO+~\cite{snoplus} collaboration.  RAT brings together the electromagnetic and hadronic physics simulation provided by GEANT4, the data storage and processing tools provided by ROOT, and parts of the scintillation and PMT simulation from GLG4sim, an open source package released by the KamLAND collaboration.  

RAT simulates the following detector effects:
\begin{itemize}
\item Propagation of primary particles, such as electrons, gamma rays, nuclear recoils, and neutrons through detector materials using GEANT4.
\item Production of extreme UV (EUV) scintillation light by charged particles in the liquid argon.  This includes the energy and particle dependence measured in \cite{microclean2008}.
\item Propagation of individual EUV photons and wavelength-shifted photons through the detector with GEANT4, including both bulk and surface optical processes of liquid argon, TPB, glass, and metal surfaces.
\item Detection of photons at PMTs and the production of realistic pulses including time, charge and shape variations, as well as pre-pulsing, late-pulsing, double-pulsing, and after-pulsing.
\item Detector triggering, waveform digitization, zero suppression, and data packing for readout.
\end{itemize}
The output of the simulation stage is formatted identically to real detector data, allowing for the development of analysis algorithms that can directly operate on both simulated and real events.

A detailed geometry of the DEAP-1 detector in the configuration described above was used to generate events simulating the \na calibration data.  The source configuration is modeled on the physical location of the source and trigger PMT with 511~keV gamma rays produced towards the center of the detector in a 9$^{\circ}$ cone coincident with a directionally uncorrelated 1274~keV gamma ray.

The distributions of the \fp, \rp, and \lr test statistics for simulated \na calibration data is shown in Figure~\ref{fig:deap1_2d} along with the mean profile for simulated nuclear recoils for comparison.  The simulated \na test statistic distributions broadly reproduce the basic shape of the distributions found in the data, although with limited statistics.  In particular, the nuclear recoil-like events begin to emerge at positive values of \lr as in the calibration data.

\section{The MiniCLEAN Detector}
\label{sec:miniclean}

\begin{figure}[htbp]
\begin{center}
\includegraphics[width=1.0\columnwidth]{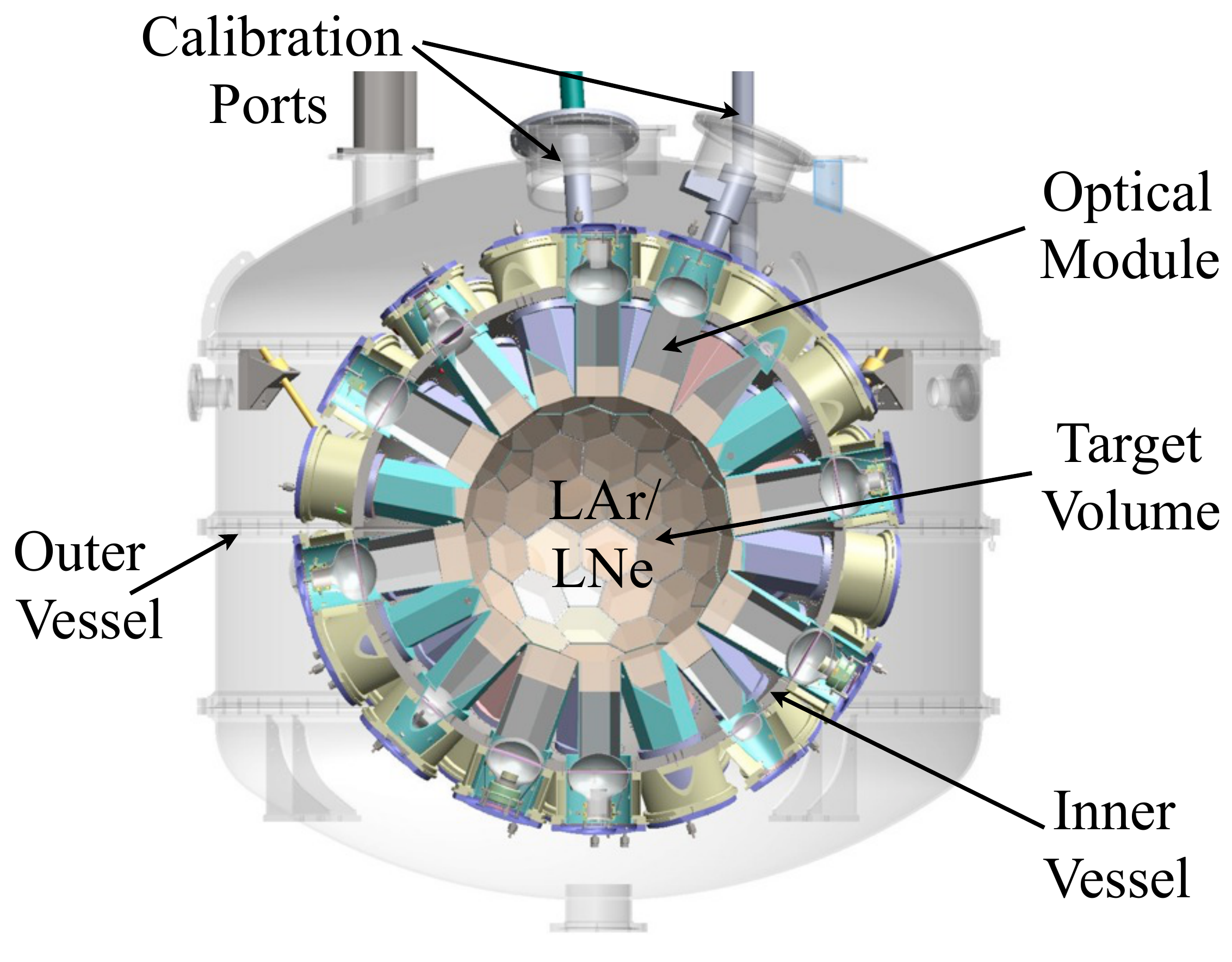}
\caption{\label{fig:miniclean_render}A 3D rendering of the MiniCLEAN detector.  The optical modules have TPB-coated acrylic at the innermost radius and PMTs at the outermost radius.}
\end{center}
\end{figure}

The MiniCLEAN detector~\cite{miniclean,minicleantaup}, shown in Figure~\ref{fig:miniclean_render}, is a cryogenic scintillation detector capable of operating with liquid argon or liquid neon.  The 500~kg central volume is surrounded by a spherical array of 92 light guides each viewed by a Hamamatsu Photonics cryogenic 8" R5912-02-MOD PMT.  The conceptually simple, scalable design with maximal PMT coverage is similar to the approaches taken by other large dark matter and neutrino detectors (for example DEAP-3600~\cite{deap3600}, SNO+~\cite{snoplus}, KamLAND~\cite{kamland}, etc).  Unlike the smaller two-PMT DEAP-1, MiniCLEAN's 92 PMTs allow 3D position reconstruction. With a complete 3D vertex, MiniCLEAN is able to apply a strict fiducial volume cut, significantly reducing backgrounds from the TPB surfaces and PMTs.  The large array of PMTs, relative to DEAP-1, with maximal coverage also reduces the pileup of scintillation photons and leads to a higher expected light yield in the MiniCLEAN design.  Both effects contribute to improved particle identification as described in Section~\ref{sec:fp_model}.

Each of MiniCLEAN's PMTs is fully immersed in the liquid and mounted in a cassette assembly that optically couples the PMT to a central target volume with a reflective tube capped by a 10~cm thick acrylic plug.  The acrylic plug acts as a neutron shield and a supporting substrate for a 2~$\mu$m layer of TPB that directly faces the argon volume to convert the extreme UV scintillation photons to the visible.  The TPB-coated acrylic blocks are suspended inside the detector by the cassette bodies to create a polyhedral approximation of a sphere at an approximate radius of 43.7~cm with 97\% of the surface area covered with TPB.  When filled with liquid argon, the total target mass is 500~kg, and after a nominal fiducial volume cut of 29.5~cm, the fiducial mass is 150~kg.

The data acquisition system is similar to the DEAP-1 design, employing 13 CAEN V1720 250~MHz, 12-bit digitizers that compress 16~$\mu$s waveforms into non-contiguous blocks of samples by applying zero-suppression.  Each block of samples contains some number of pre- and post-samples that extend beyond the samples exceeding the zero-suppression threshold.  In a typical configuration, the minimum size of a block containing at least one photoelectron pulse is 32 samples.  The CAEN V1720 digitizers have been observed in the lab to have a noise RMS of 0.41~mV per sample when connected to a full length coaxial cable and PMT.

\section{Simulation of MiniCLEAN}
\label{sec:sim}

\begin{figure}
\begin{center}
\includegraphics[width=1.0\columnwidth]{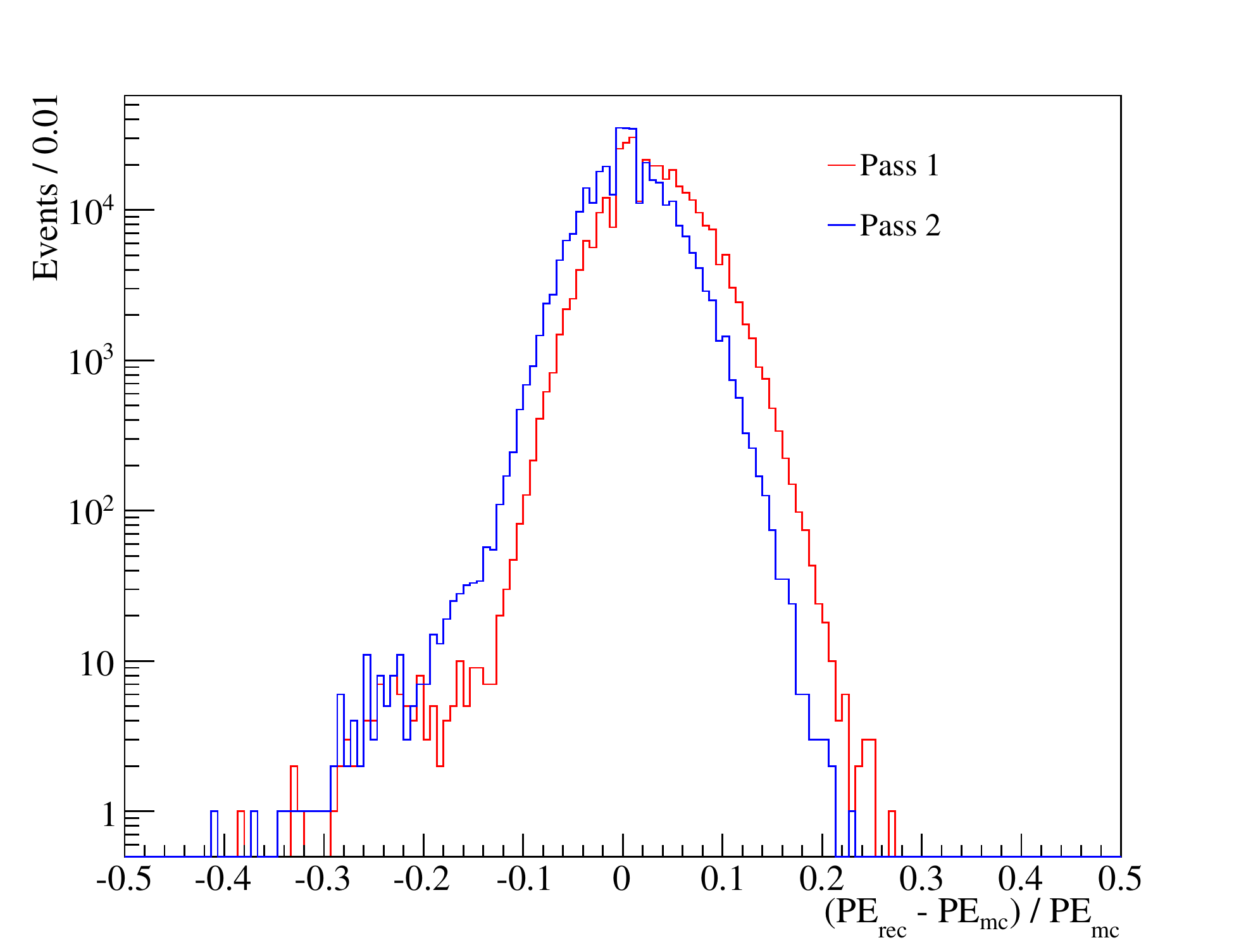}
\caption{\label{fig:bayes_2pass}The fractional error in estimation of the number of photoelectrons in simulated electron events in the MiniCLEAN detector between 75 and 100 PE.  The second pass of Bayesian photoelectron counting, which replaces the bootstrapped Bayesian prior with a prior derived from the reconstructed position and energy of the event, reduces the energy bias from 3.0\% to 0.5\%, reduces the energy resolution from 4.1\% to 3.9\%, and reduces the RMS of the \lr distribution by 5\%.}
\end{center}
\end{figure}

A sample of 84 million \artn decays were simulated and processed through the entire analysis sequence as described in Section~\ref{sec:spe_counting}.  The events have been limited to be within the fiducial radius, where the energy scale has no radial dependence.  For 75-100 PE (12.5-16.7 keVee) events, Figure~\ref{fig:bayes_2pass} shows the fractional error in the number of photoelectrons estimated using the Bayesian photoelectron counter on the first pass with bootstrapped priors compared to the second pass where the priors, $\mu$ from Equation~\ref{eq:p_n}, are updated using position reconstruction.  Although the updated Bayesian priors can induce small chennel-to-channel correlations, the second pass of the algorithm is an overall improvement.  The second pass of Bayesian photoelectron counting reduces the energy bias from 3.0\% to 0.5\% and improves the energy resolution from 4.1\% to 3.9\%.

\begin{figure}
\begin{center}
\includegraphics[width=1.0\columnwidth]{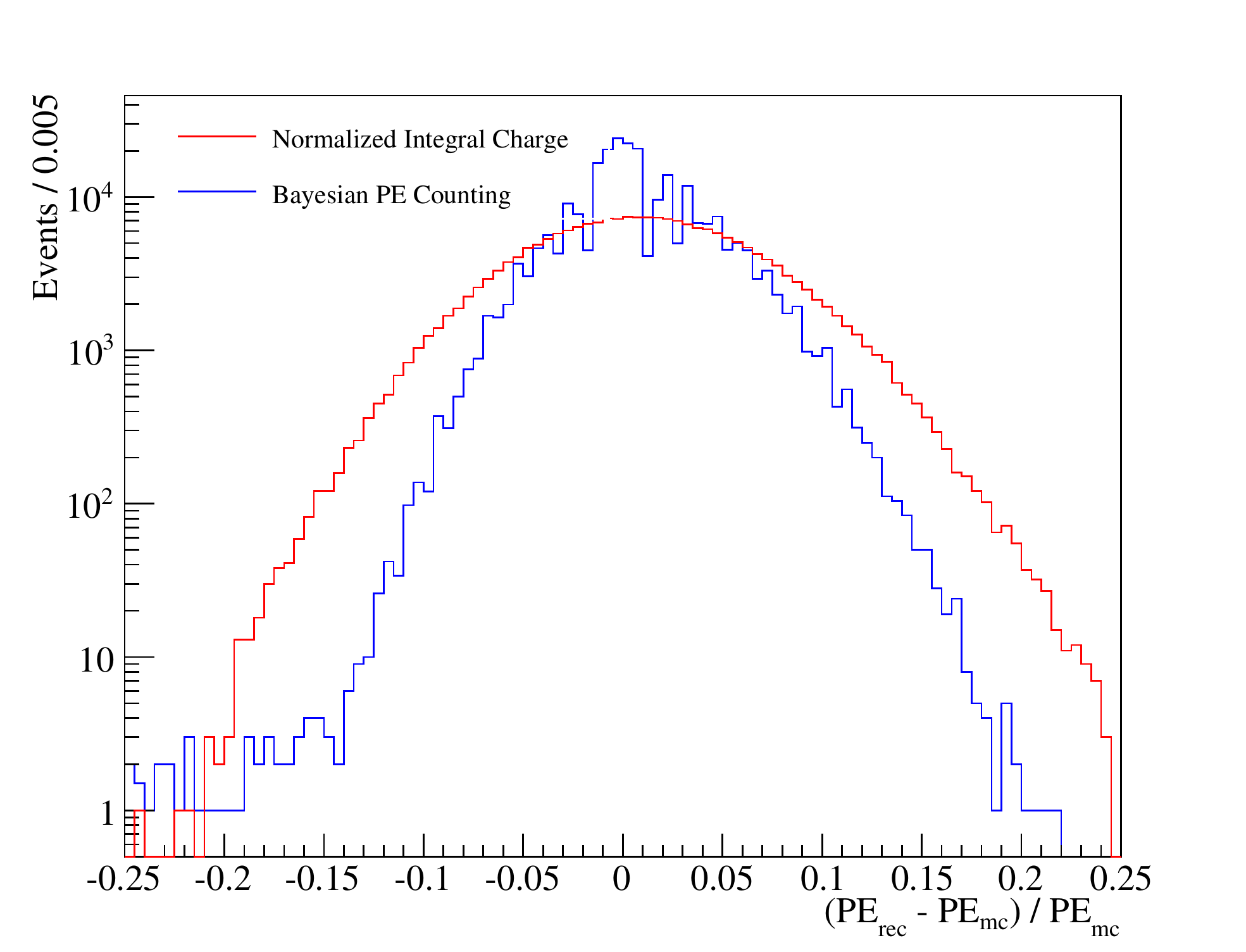}
\caption{\label{fig:eres_miniclean}The ratio of estimated number of photoelectrons divided by the true number of photoelectrons for simulated \artn decays in the MiniCLEAN.  Only events between 75 and 100 PE, and radius less than 295 mm are shown.}
\end{center}
\end{figure}

Figure~\ref{fig:eres_miniclean} compares the apparent event energy from normalized charge integration, where the number of photoelectrons is estimated from the total charge divided by the mean single photoelectron charge, to the number of photoelectrons estimated using the two-pass Bayesian photoelectron counter.  Since events are restricted to a region where the energy scale has no radial dependence, the improvement in photoelectron counting resolution is a direct measure of the improvement of the energy resolution.  The energy resolution is reduced from 4.3\% with the charge integration method to 3.1\% with the Bayesian photoelectron counting.

The improved energy resolution offered by the Bayesian photoelectron counting reduces the number of low energy background events entering the energy region of interest.  Lowering of the threshold with reduced background in the presence of a WIMP-like signal can dramatically improve the sensitity to signal which increases rapidly at low energies.

\begin{figure*}
\begin{center}
\includegraphics[width=1.0\textwidth]{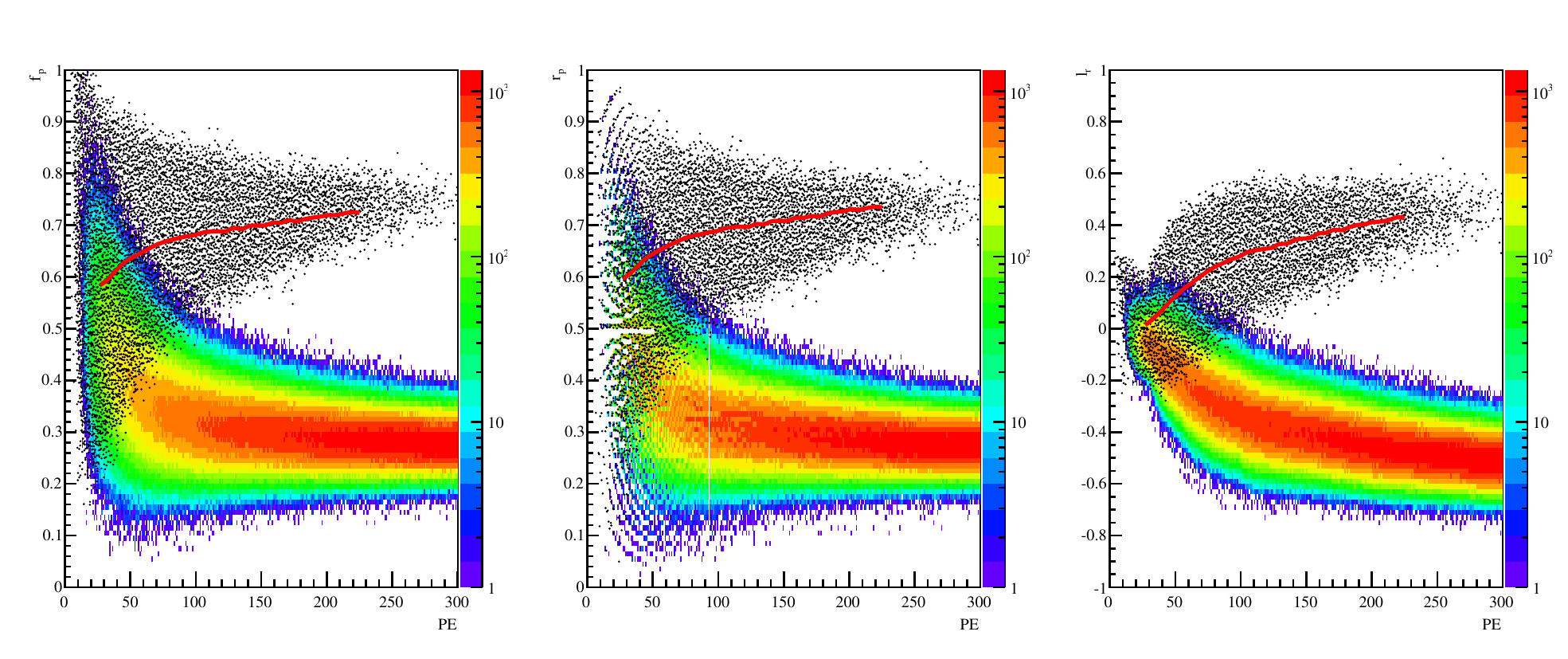}
\caption{\label{fig:mini_2d}Distribution of \fp (left panel), \rp (center panel), and \lr (right panel) test statistics for simulated electronic (color scale) and nuclear (black points) recoils as a function of reconstructed number of photoelectrons in the MiniCLEAN detector.  The solid red line in each panel indicates the mean profile for nuclear recoil.}
\end{center}
\end{figure*}

\begin{figure}
\begin{center}
\includegraphics[width=1.0\columnwidth]{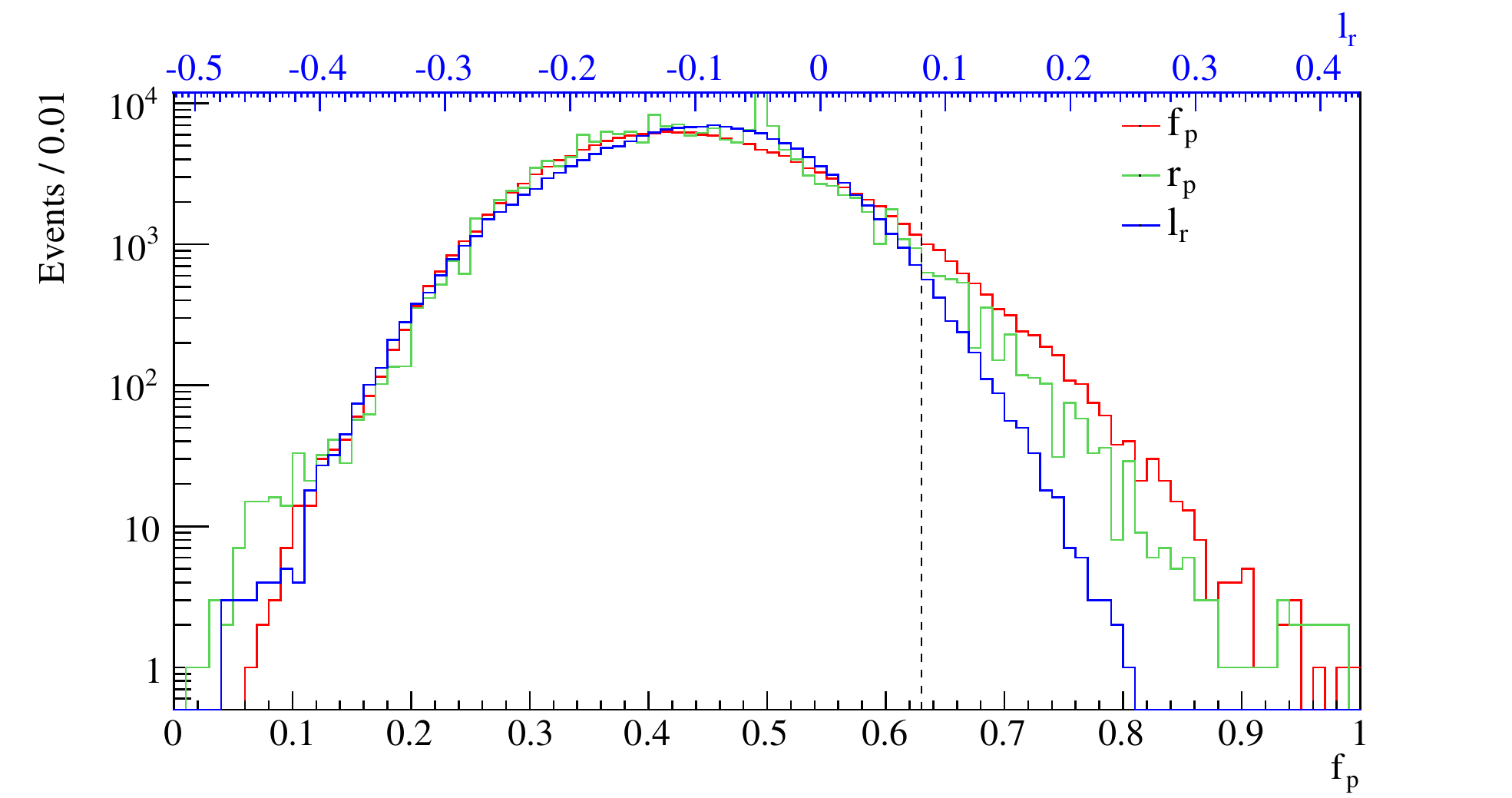}
\caption{\label{fig:miniclean_1d_compare}Distribution of \fp, \rp and \lr test statistics for electronic recoils for simulated MiniCLEAN events at 40 PE.  The vertical dashed line indicates 50\% nuclear recoil acceptance at 6.67~keVee.  The \lr values have been linearly transformed such that the median values for the electron and nuclear recoil distributions match those for \fp.}
\end{center}
\end{figure}

The 2D distributions of the \fp, \rp, and \lr test statistics as a function of reconstructed number of photoelectrons are shown in Figure~\ref{fig:mini_2d} for simulated events in the MiniCLEAN detector.  The \lr test statistic significantly tightens up the distributions of electrons and nuclear recoils at very low numbers of photoelectrons, which is precisely where a WIMP signal will show increasing rate.  We do not have enough CPU resources to fully simulate the number of events required to estimate the background leakage for each of these test statistics directly, as we did with the fast simulation in Section~\ref{sec:fp_model}.  However, we can linearly transform \lr into the same [0,1] range as \fp and \rp so the median \lr value for electrons is the same as the median \fp, and the same is true for nuclear recoil distributions.  This allows the tails of the distributions to be compared more easily, as is shown in Figure~\ref{fig:miniclean_1d_compare}.  Electrons with only 40~PE (6.67~keVee as in Figure~\ref{fig:deap1_1d_compare} for DEAP-1 given the respective light yields) show a significantly smaller tail in the \lr distribution compared to \fp and \rp.  Above \fp of 0.7, the \lr tail is reduced by a factor of $7.2\pm0.2\;\mathrm{(stat)}$ compared to the \fp distribution.

\section{Conclusion}

Many current and future neutrino and dark matter experiments depend on the detection of scintillation light with PMTs.  The ability to both count and identify the times of individual photoelectrons in a PMT waveform directly impacts energy resolution and particle identification with timing.  When scintillation light with both short and long time constants is present, simple strategies for photoelectron counting will alternately introduce biases or large variance in different situations.  We have resolved this difficulty by showing how to apply a Bayesian prior assumption about the intensity and time structure of the light to the counting process.  By iterating the procedure, a crude, bootstrapped prior can be refined with the addition of reconstructed observables, such as event energy and position.

Although the canonical prompt-fraction, \fp, particle identification parameter often offers excellent discrimination, realistic detector effects can degrade the discrimination power significantly with the extent depending on the scintillator and detector properties.  Using the times output by the Bayesian photoelectron counting procedure, improved test statistics, such as \rp and \lr, can be constructed to better utilize the known time structure of the scintillation light.  Although this is applicable to a wide range of scintillator detectors, here we have demonstrated the improved particle identification test statistics in the DEAP-1 \na calibration data and MiniCLEAN \artn simulation.  With improved energy resolution and improved particle identification test statistics, scintillator experiments looking for rare events can improve sensitivity to the signal of interest through the reduction of backgrounds and lowering of the energy threshold.

\section*{Acknowledgments}

This work has in part been supported by the United States Department of Energy, Office of High Energy Physics.

Support for DEAP-1 has been provided by the Canadian Foundation for Innovation and the Natural Sciences and Engineering Research Council.  The High Performance Computing Virtual Laboratory (HPCVL) has provided us with CPU time, data storage, and support.  We would also like to thank the SNOLAB staff for on-site support.  The work of our co-op and summer students, including Christopher Stanford who operated DEAP-1 during the data-taking in this paper, is gratefully acknowledged.

Contributions by staff of NIST, an agency of the US government, are not subject to copyright in the US.

\end{document}